\documentclass[%
 aip,
amsmath,amssymb,
preprint,%
]{revtex4-1}

\usepackage{graphicx}
\usepackage{dcolumn}
\usepackage{bm}
\usepackage[mathlines]{lineno}

\usepackage[utf8]{inputenc}
\usepackage[T1]{fontenc}
\usepackage{mathptmx}
\usepackage{etoolbox}
\usepackage{xcolor,amsmath}
\usepackage{booktabs}

\newcommand{\of}[1]{\left( #1 \right)}

\newcommand{\sqbs}[1]{\left[ #1 \right]}

\newcommand{\E}[1]{\mathbb{E}\sqbs{#1}}

\makeatletter
\def\@email#1#2{%
 \endgroup
 \patchcmd{\titleblock@produce}
  {\frontmatter@RRAPformat}
  {\frontmatter@RRAPformat{\produce@RRAP{*#1\href{mailto:#2}{#2}}}\frontmatter@RRAPformat}
  {}{}
}%
\makeatother
\begin{document}


\title{Exploring the Dynamics of Lotka-Volterra Systems: Efficiency, Extinction Order, and Predictive Machine Learning}
\author{Sepideh Vafaie}
\affiliation{Department of Earth and Environmental Studies, Montclair State University, 1 Normal Avenue, Montclair, NJ 07043, USA}
\affiliation{School of Computing, Montclair State University, 1 Normal Avenue, Montclair, NJ 07043, USA}
\altaffiliation{Author to whom correspondence should be addressed: \href{mailto:vafaies1@montclair.edu}{vafaies1@montclair.edu}}
 
\author{Deepak Bal}%
\affiliation{Department of
     Mathematics, Montclair State University, 1 Normal Avenue, Montclair, NJ 07043, USA\\
}%

\author{Michael A.S. Thorne}
\affiliation{%
British Antarctic Survey, High Cross,
Madingley Road, Cambridge, CB3 0ET, UK
}%
\author{Eric Forgoston}
\affiliation{School of Computing, Montclair State University, 1 Normal Avenue, Montclair, NJ 07043, USA}
\date{\today}

\begin{abstract}
For years, a main focus of ecological research has been to better understand the complex dynamical interactions between species which comprise food webs.
Using the connectance properties of a widely explored synthetic food web called the cascade model, we explore the behavior  of dynamics on Lotka-Volterra ecological systems. We show how trophic efficiency, a staple assumption in mathematical ecology, produces systems which are not persistent. With clustering analysis we show how straightforward inequalities of the summed values of the birth, death, self-regulation and interaction strengths provide insight into which food webs are more enduring or stable. Through these simplified summed values, we develop a random forest model and a neural network model, both of which are able to predict the number of extinctions that would occur without the need to simulate the dynamics. To conclude, we highlight the variable that plays the dominant role in determining the order in which species go extinct.  
\end{abstract}

\maketitle

\begin{quotation}
A food web describes the complex dynamical relationships between species or groups of species. While these relationship interactions can be mathematically modelled using the Lotka-Volterra equations, ecologists often analyze a food web system via the community matrix, i.e. the Jacobian of the nonlinear system evaluated at equilibrium. To better understand the interaction dynamics, we instead consider the full, nonlinear dynamics associated with a type of synthetic food web (the cascade model), and examine various properties associated with the extinction dynamics and species persistence.
Even relatively small food webs consist of hundreds of interactions which make analysis of food webs extremely complicated. By reducing the many interactions in the governing equations to the absolute value of the sum of birth, death, self-regulation, predation, and prey rates we have been 
able to determine certain properties related to persistence in food webs. Applying machine learning approaches trained on the simplified input of the five rate sums leads to accurate predictive models of the persistence of a given system without the need to simulate the dynamics. Consideration of the simplified rate sums also enables us to explore the internal processes involved in the extinction of species during
the unfolding dynamics of a food web ecosystem. 

\end{quotation}

\section{\label{sec:level1}Introduction}
A food web can be described as a complex network of interactions between consumers and resources involving organisms, populations, or trophic units \citep{ winemiller1996food}. Elton represented food webs as diagrams showing the energy flow within the respective consumer or resource groups throughout the system \citep{elton1927}. Such energy flow interactions typically represent predator-prey, competitive, and mutualistic consumer/resource relations \citep{ pimm1982food}, where the edge connections illustrate which group or species are consumed. Other types of food webs {include} topological food webs and functional food webs. A topological food web emphasizes the feeding relationships among species, which can be observed or estimated. A functional food web identifies the species which are most important in maintaining the integrity of community composition and structure \citep{paine1980,winemiller1996food}.

Mathematical ecology primarily considers food webs as a fundamental unit which tends to exhibit emergent characteristics that stem from their individual components \citep{ hall1993food}. Experimentally derived food webs can help identify the species and feeding connections that have the most impact on population and community dynamics. However, selecting which species and interactions to manipulate in experiments in large, complex systems can be subjective and is often based on incomplete knowledge of the food web. Additionally, this approach faces challenges such as controlling variables, choosing the right time frame and spatial scale for experiments, choosing the initial conditions, and determining how to isolate specific sub-systems, all of which can affect outcomes \citep{bradley1983complex, bender1984perturbation, yodzis1988indeterminacy}. Mathematical ecologists have often drawn on synthetic systems to explore the more general resultant properties of community trophic networks, as problematic as such abstraction can be in its own right. 

One example of a synthetic food web that has been used to explore topological properties of systems is the cascade model \citep{cohen1985stochastic}. The cascade model, which is structurally hierarchical, describes the organization of a food web and the topology of the interactions, but it does not account for the dynamics of the system. 
However, to fully understand ecological communities, it is crucial to consider both the population dynamics (birth, death, self-regulation, and predator-prey interactions) and trophic structure.  

Incorporating Lotka-Volterra dynamics \citep{lotka1920,lotka1926,volterra1926} into a structural food web allows one to qualitatively and quantitatively predict how a population of interacting species will behave over the long term. The combination of Lotka-Volterra dynamics with a casacde food web leads to the Lotka-Volterra cascade model (LVCM) \citep{cohen1985stochastic} which enables one to explore how certain topological constraints under specific distributions behave when dynamics are introduced.

One of the big challenges in ecology and environmental science more generally is understanding the forces that shape the stability of complex systems, and conversely lead to collapse or extinction. One of the extinction scenarios that can represent a natural ecosystem's response to a realistic extinction sequence is derived from the study provided by the International Union for Conservation of Nature (IUCN)\citep{de2011serengeti}. According to this scenario, the initial wave of extinctions primarily affects large-bodied species, including predators and mega-herbivores \citep{ebenman2011response}. The likelihood of further extinctions occurring after the initial loss in food webs depends on the number of species within each functional group. Specifically, the risk associated with a particular number of species per functional group is contingent on the type of species that is removed. This risk is at its highest when an autotroph is the first to be lost and is at its lowest when a top predator is the initial casualty \citep{borrvall2000biodiversity}. The impact of primary extinctions was assessed in terrestrial and aquatic ecosystems, measuring robustness in relation to secondary extinctions \citep{dunne2002network}. A novel approach for quantifying secondary extinctions following species loss reveals that, in a deterministic context, communities with a greater number of species within trophic levels tend to preserve a higher proportion of species \citep{ebenman2004community}.

Previous studies have elucidated some of the causes behind primary and secondary extinctions, which can ultimately result in the loss of species within food webs and, in the worst-case scenario, even lead to the collapse of entire food webs \citep{dunne2002network, ebenman2004community}. These studies, however, did not explicitly include the nonlinear dynamics inherent in the system. In the late 1990s, some researchers began emphasizing nonlinear modelling, but these works generally considered the dynamics of small, specialized food webs (e.g., food chains, food chains with omnivory) \citep{mccann1997,mccann1998} or small food webs containing no more than ten species \citep{williams2004stabilization}. Later work \citep{martinez2006diversity,dominguez2019} considered larger food webs with dynamics given by a bioenergetic consumer-resource model \citep{yodzis1992}. While these studies provided insight into the complexity-stability relationship, there has been a lack of research focusing on understanding the process of species extinction through mathematical models that simulate the natural dynamics of species loss within dynamic ecosystems. In our study, we address this gap by employing the LVCM to explore the loss of species. We apply deterministic dynamics, with rates chosen randomly from a specified distribution, to a variety of randomly constructed cascade food webs to investigate this phenomenon comprehensively. 

Ecologists often use community matrices to explore general ecosystem behaviour. However, the community matrix, a Jacobian matrix associated with a linearized dynamical system (often a form of the Lotka-Volterra equations), necessarily assumes the system is at equilibrium. Moreover, the community matrix ignores the nonlinearities of the Lotka-Volterra system.  Notably, when the system possesses self regulation, such as is found in ecological systems with limited resources or negative feedback mechanisms the system can always be made stable by increasing the amount of self-regulation \citep{thorne2021matrix}. This simple fact can easily be deduced via the Gershgorin circle theorem \citep{gershgorin1931uber,horn2012}. But to fully understand the dynamics of synthetic or real food webs, one should comprehensively investigate how the full Lotka-Volterra system governs the behaviour of ecosystems. Such a study should include the role of the birth, death, self-regulation, and predator-prey rates along with the effect of the trophic efficiency parameter on species extinction.  

Sections \ref{sec:cm} and \ref{sec:LVD} contain details of the cascade model and the LVCM respectively. Section \ref{sec:results} contains the analytical and numerical results, including an investigation of the impact of trophic efficiency on extinction in Section \ref{sec:efficiency} and of the effect of the rates on persistence in Section \ref{sec:persistence}. In the latter section, we showed how a reduced amount of information involving five absolute rate sums, rather than the hundreds of values of individual rates, can explain when persistence can be attained.  In {Sections \ref{sec:prediction-c} and}  \ref{sec:prediction} we {respectively} use these five sums as input to {a random forest model and a} minimal neural network, {both of} which enable the prediction of how many species will go extinct with very high accuracy. In Section \ref{sec:order} we show which dynamics determine the order of species extinction. Lastly, we close with some comments in Section \ref{sec:conc}. 
\section{\label{sec:cm}Cascade Model}
A synthetic foodweb construction, called the cascade model
 \citep{cohen1985stochastic}, was suggested roughly forty years ago as a  framework to generate the connectence properties of ecological communities, with the intention that they reflect the accurate topology of systems found in nature. The construction of a cascade model is as follows. Consider a community with $S$ species where each species is assigned a unique integer label from $1$ to $S$. It is important to note that the structure of a cascade food web is hierarchical, and any given species is not able to prey upon species which are denoted by higher number values. 
 
The cascade model of a food web can be visually represented using a random directed graph with $n$ vertices. Each vertex corresponds to a species, and $P_w(i,j)$ determines the likelihood of an arrow from species $i$ to species $j$. If $1\leq i < j\leq S$, then $P_w(i,j)=0$. On the other hand, if $i>j$, then $P_w(i,j)=c/S$, where $c$ is essentially the average total degree (number of in and out edges) of each species node.  In short, if the vertex $i$ has a higher label than $j$, there is no link from species $j$ to species $i$ with probability of one. Conversely, there exists a link from species $i$ to species $j$ with  probability of $c/S$. 
In this work, we follow Cohen and Newman \citep{cohen1985stochastic}  and let $c=3.72$. With this value of $c$, the model was considered to have reasonable alignment between the observed proportions in real food webs and the predicted proportions generated by the cascade model for various types of links, including those between basal-intermediate species and basal-top species \citep{cohen1985stochastic}.

\begin{figure}[t!]
\centering
\includegraphics[scale=0.45]{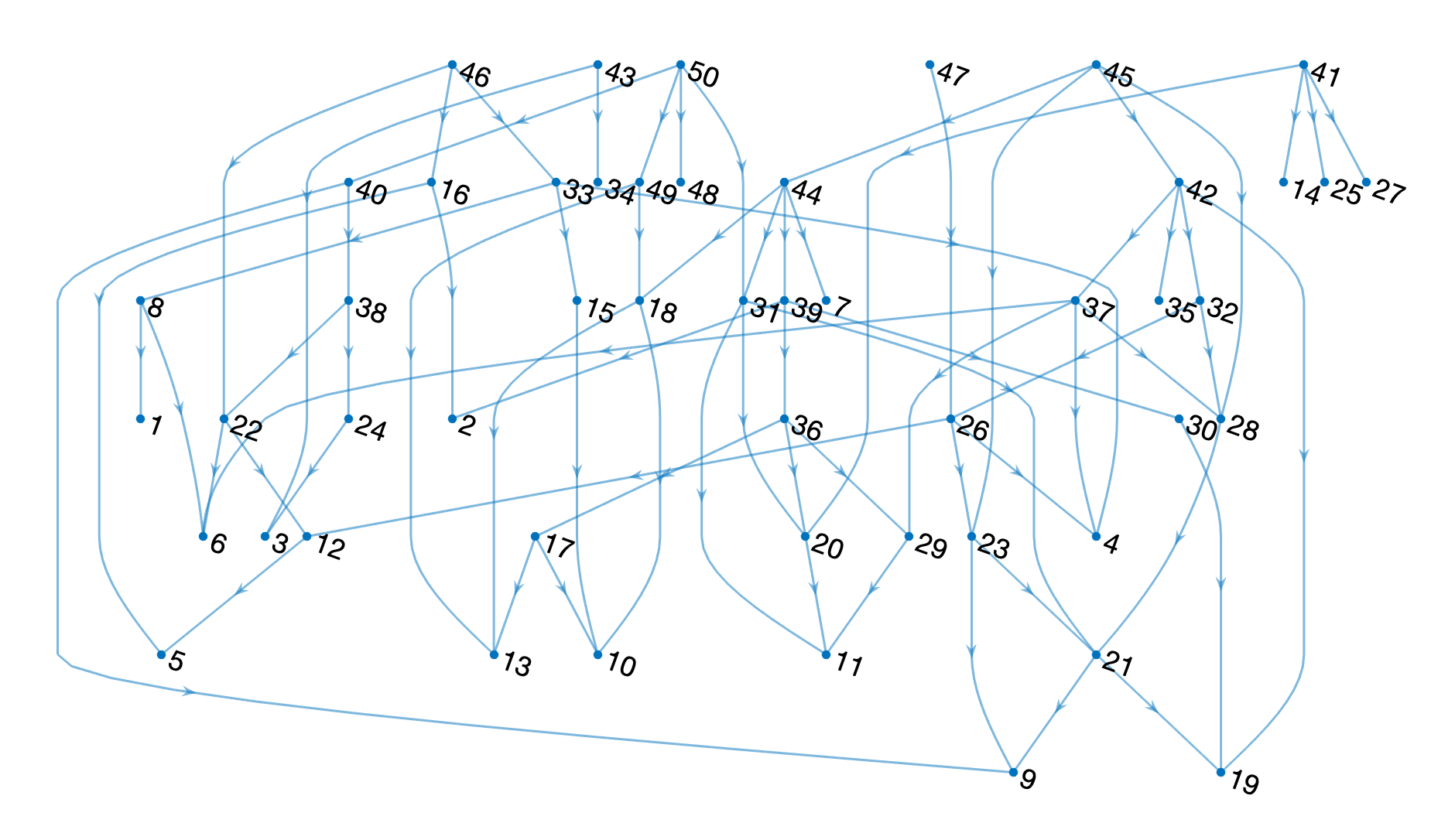}
\caption{A random realization of a cascade model food web with 50 species. Nodes represent uniquely labeled species, and arrows point from predators to prey.}
\label{fig:cascade}
\end{figure}

Figure \ref{fig:cascade} illustrates a realization of the cascade model food web with 50 species. Each node is indicative of a distinct species, labeled uniquely. The arrows depict the direction from predators to prey. Within our investigation, we categorize species into two types: basal and non-basal. Basal species serve only as prey for other species in the food web, and do not predate upon other species. Non-basal species take on the roles of both prey and predator, unless the species is a top-predator, in which case it only predates upon other species. For instance, in Figure \ref{fig:cascade}, species 46 is classified as non-basal and preys upon species 16, 22, and  33. Species 19 is considered basal, as it does not prey upon any other species. 
It is noteworthy that the construction of the cascade model came at a na\"{i}ve time in mathematical ecology when it was considered that the topological properties of food webs were their defining criteria \citep{may2019stability}, with little emphasis placed on the importance of interaction strength. Similarly, the dynamics that may have led to the equilibrium of which the topology is an endpoint was little explored.
\section{\label{sec:LVD}Lotka-Volterra Cascade Model (LVCM)}
The dynamics of predator-prey interactions are modeled using the Lotka-Volterra system of equations, which are given as
 \begin{equation}
    \frac{dX_i}{dt} = X_i\Bigg(b_i + \sum_{j} a_{ij}X_j\Bigg),
\quad    i=1,...,S, \label{equ:L-V}
\end{equation}
with $a_{ji}=-ea_{ij}$, for $i\neq j$. In Equation (\ref{equ:L-V}), 
$X_i$ represents the population density of species $i$, $b_i$ denotes the natural growth rate of species $i$, $a_{ij}$ indicates the interaction rate between species $i$ and $j$, and parameter $S$ represents the number of species in the food web.

For the purposes of the study, we have chosen the initial value of $X_i$ to be uniformly distributed on the interval $(0, 1)$ for each species. For basal species, $b_i$ represents the birth rate and is uniformly drawn from the interval $(0, r)$, where $r>0$. The specific value of $r$ is specified throughout Section \ref{sec:results}. For nonbasal species, $b_i$ represents the death rate, and it is uniformly distributed on the interval $(-r, 0)$. In the interaction matrix $\bf{A}$, the diagonal elements $a_{ii}$ correspond to self-regulation within a species, and the value of each $a_{ii}$ term is drawn uniformly from $(-r,0)$. The elements $a_{ij}$ residing above the diagonal of $\bf{A}$ represent the interaction rate between predator species $j$ feeding on prey species $i$, and are chosen uniformly from $(-r,0)$. The values for $a_{ji}$  are determined by $a_{ji}=-ea_{ij}$, where $e$ represents the trophic efficiency, which signifies how efficiently predator growth is initiated by prey consumption. In ecological modeling, this parameter helps describe how effectively an organism utilizes the energy it obtains from consuming resources for the purpose of reproduction. Biologically, the value of the trophic efficiency parameter, $e$, ranges between zero and one, where zero signifies no trophic efficiency in converting consumption into reproduction, and one indicates maximum trophic efficiency \citep{ginzburg1998assuming}.

In this article, we consider the Lotka-Volterra cascade model (LVCM) \citep{1cohen1990stochastic}. This hybrid model allows one to capture the dynamics of species populations along with trophic structure in an ecological community. The construction of the LVCM consists of two steps. First, we use the cascade model to generate a food web. Second, Lotka-Volterra dynamics according to Equation (\ref{equ:L-V}) are applied to the cascade food web generated in the first step.
When employing this process, it is commonly observed that the inclusion of dynamics into the cascade food web leads to the occurrence of numerous extinction events within the ecosystem. Eventually, the system evolves to a new stable food web at equilibrium. {Similar extinction behaviour has been seen in niche model food webs with consumer-resource dynamics \citep{dominguez2019}.}

\begin{figure}[t!]
\begin{center}
\includegraphics[scale=0.45]{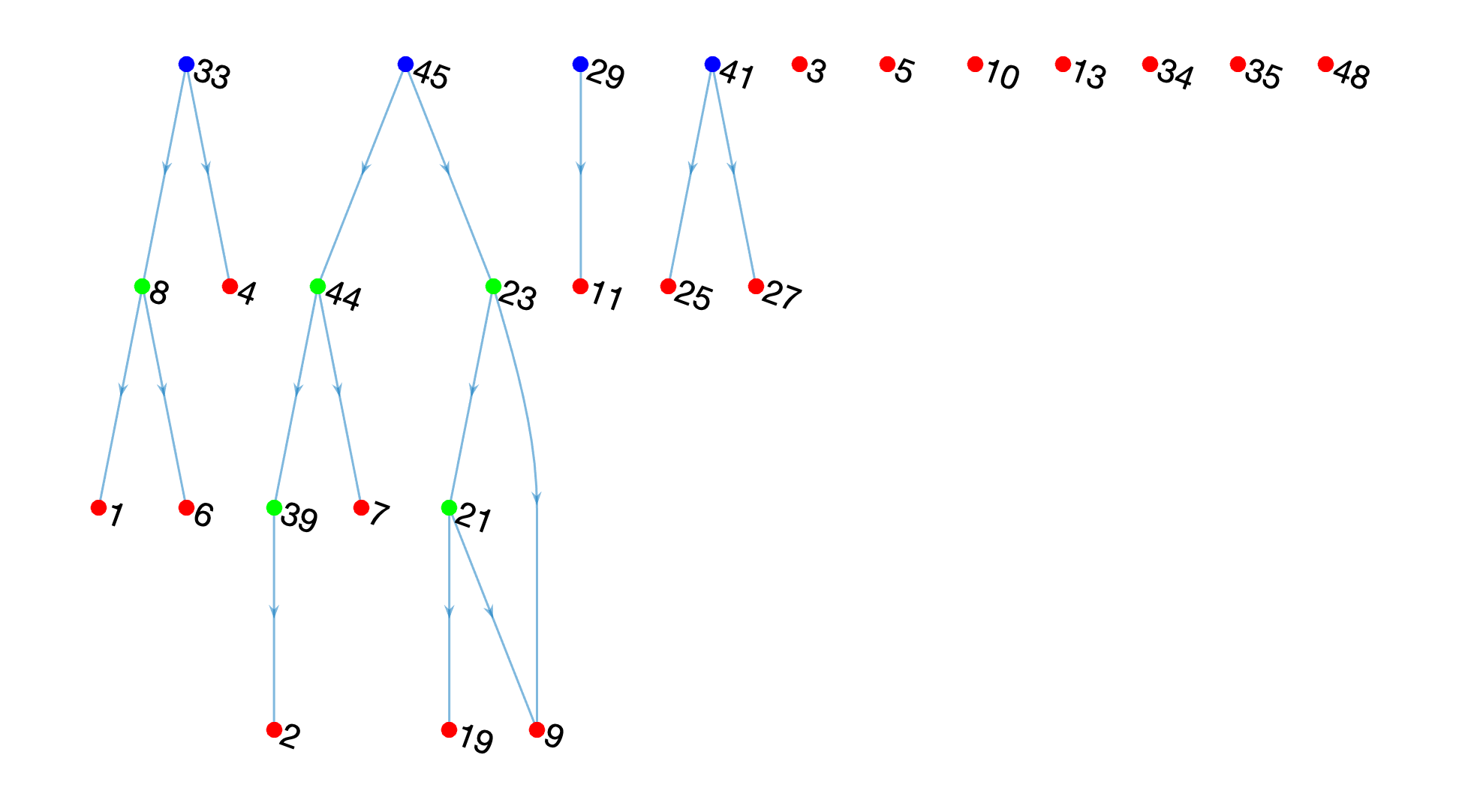}
\caption{The resulting stable food web after applying Lotka-Volterra dynamics (Equation (\ref{equ:L-V})) to the cascade food web depicted in Figure \ref{fig:cascade}. Although the original cascade food web had 50 species, the dynamics induced numerous species extinctions. In this reduced, stable food web, only 26 of the original species have survived. The stable food web includes surviving basal species shown in red, top predators in blue, and intermediate predators in green. This extinction phenomena is general, and can be seen for different cascade food webs of different sizes and for different parameter values in the dynamics.}
\label{fig:cascade.L-v}
\end{center}
\end{figure}

As an example, when dynamics are introduced to the food web shown in 
Figure \ref{fig:cascade}, using $e=0.8$ and $r=1$, many species go extinct until eventually the system evolves to the new, stable food web given by Figure \ref{fig:cascade.L-v}. As shown in Figure \ref{fig:cascade.L-v}, the new stable food web that is generated after applying the dynamics given by Equation (\ref{equ:L-V}) is comprised  of surviving basal species shown in red, as well as top predators indicated in blue, and intermediate predators in green. It is important to note that top predators in the reduced, stable food web may not have been top predators in the original food web before dynamics were applied to the food web. 

The LVCM provides a framework for modeling real ecological communities and enables improved understanding of the intricate predator-prey relationships in a food web. By considering both the population dynamics and trophic structure, we can gain insight into the functioning of ecosystems.

\section{\label{sec:results}Results}
We have incorporated Lotka-Volterra dynamics into the synthetic cascade food web model and have investigated how the introduction of dynamics can lead to the extinction of species, thereby changing the food web's structure. Using analytical and numerical methods, we have investigated the influence of dynamical rates and predation efficiency on species persistence. Additionally, we have utilized random forest and neural network models to predict how many extinctions occur under Lotka-Volterra dynamics. Notably, both models allow for explainability of the results via the relative importance of the features used as inputs to the models. Moreover, we have derived an analytical expression to elucidate the sequence of species extinctions.
\subsection{\label{sec:efficiency}Impact of Efficiency on Extinction}
We have investigated the impact of the energy trophic efficiency parameter, $e$, on species extinction using the LVCM which was described in Section \ref{sec:LVD}. As can be seen from Equation (\ref{equ:L-V}), the trophic efficiency parameter establishes the relationship between the interaction rates $a_{ij}$ and $a_{ji}$. Biologically, the trophic efficiency parameter lies within $(0,1)$ \citep{ginzburg1998assuming}.

\begin{figure*}

\centering
\includegraphics[scale=0.41]{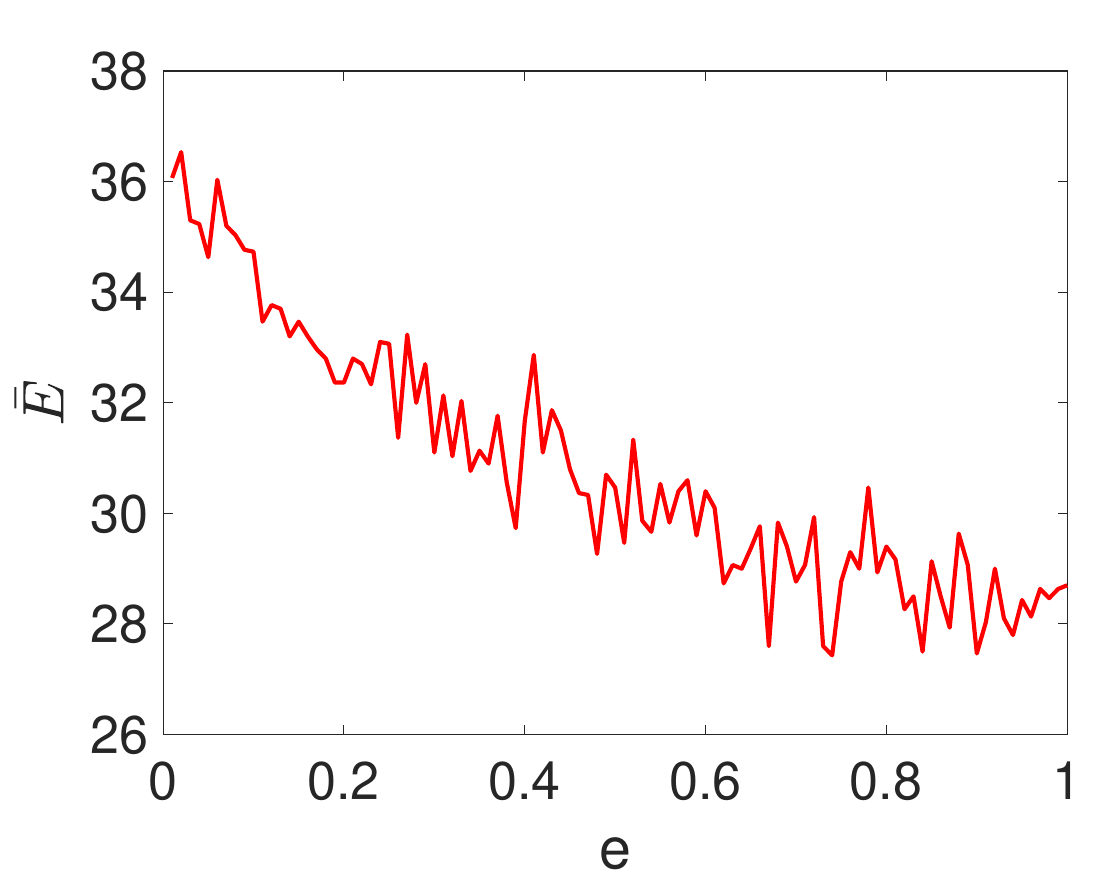}
\includegraphics[scale=0.41]{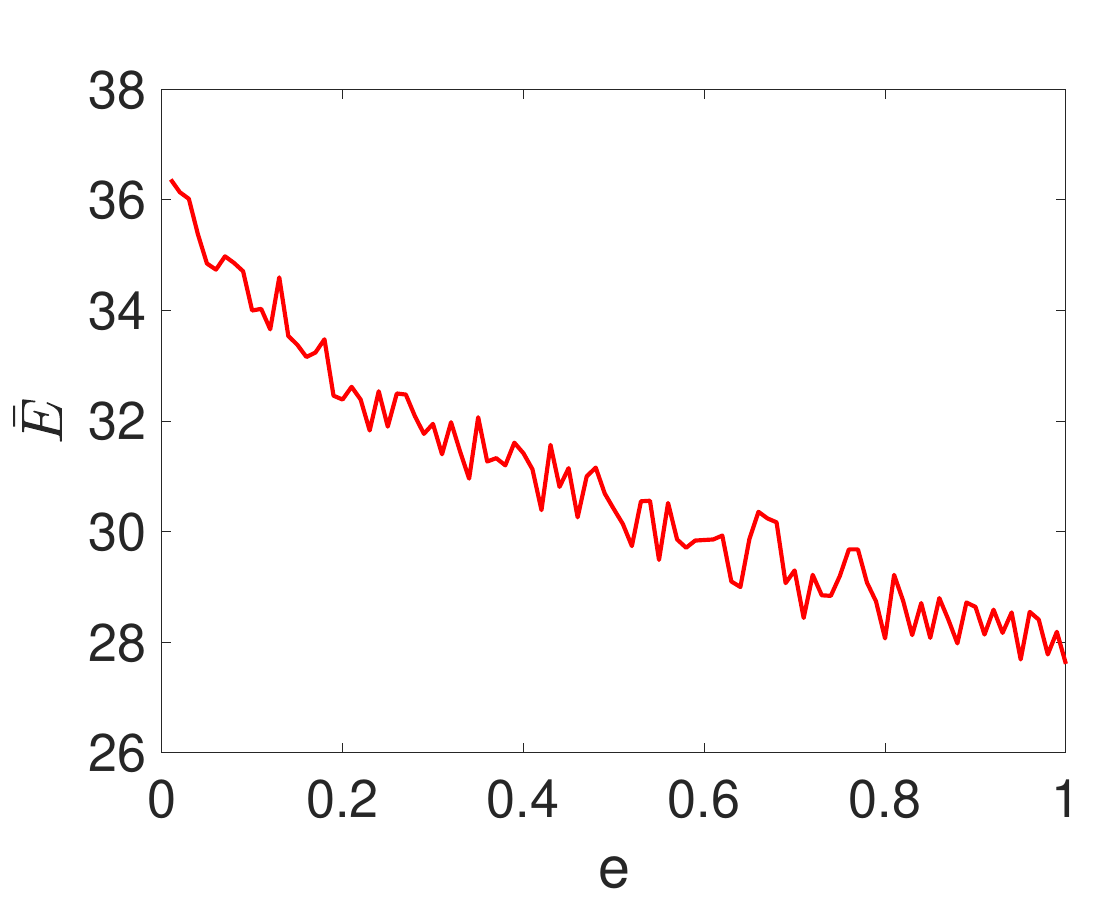}
\caption{Average number of extinctions, $\bar{E}$, as a function of efficiency, $e$, for LVCM food webs of 50 species. The average is computed for (a) 30 realizations, and (b) 100 realizations of food webs and associated Lotka-Volterra rates.}
\label{fig:efficiencie}
\end{figure*}

To explore the role which efficiency plays with regards to species extinction, we consider efficiency  values which range from 0.01 to 1.0 in increments of 0.01. For each trophic efficiency value, we generate $30$ different cascade food webs consisting of 50 species. For each food web, Lotka-Volterra dynamics (Equation (\ref{equ:L-V})) with randomly generated rates using $r=1$ as described in Section \ref{sec:LVD} are applied, and the number of species extinctions which occur is recorded.  For each  trophic efficiency value, the average number of extinct species, $\bar{E}$, across the 30 realizations is computed. Figure \ref{fig:efficiencie}(a) shows the result by plotting the average number of extinct species as a function of efficiency value. 
One can clearly see that as the efficiency increases, the number of extinct species decreases, dropping from about 36 to about 29. Even though higher efficiencies do decrease the number of extinctions, there are still far too many extinctions for the food web to persist. Indeed, even at maximum efficiency, more than half of the species in the food web are going extinct. 
If one increases the efficiency value beyond one, it is possible to achieve a further decrease in the number of extinctions, including instances of zero extinctions. However, values of efficiency greater than unity are not biological.

It is worth mentioning that the fluctuations seen in Figure \ref{fig:efficiencie}(a) arise from two main factors. First, if one were to increase the number of realizations for each simulation, one would reduce the impact of statistical fluctuations, and the resulting curve would be much smoother (Figure \ref{fig:efficiencie}(b)). This does not affect the overall trend described above. Second, efficiency is not the sole parameter governing the number of extinct species. Many other factors, which we discuss below, also play a role. Taken together, the complex dynamics can lead to fluctuations, but again, do not affect the overall trend. 

The results above indicate that constraining the upper triangular values of $\bf{A}$ to be linked to the lower triangular values of $\bf{A}$ through the efficiency variable, $e$, ensures that the systems are unlikely to be viable. Therefore, for much of the following results, we remove the efficiency link between the upper and lower triangular values, letting their distributions be independent of each other.

We carried out a similar analysis on another popular synthetic food web model, the niche model \citep{williams2000}, and found that efficiency acts in an equivalent manner in causing mass extinction of the system.

\subsection{\label{sec:persistence}Effect of Rates on Persistence}
We now consider the effect on persistence of the different dynamical rates found in the Lotka-Volterra equations. At a foundational level, for any food web, if there is not enough basal species biomass available for consumption by species at higher trophic levels, then some non-basal species will go extinct. To ensure persistence of the food web, the growth rates of the basal species must be high enough to allow the non-basal species to thrive. Similarly, if the death rate of a non-basal species is too high to be offset by gains from prey consumption, the species will go extinct. 

To fully understand the role the rates play on the food web dynamics, we generated 400,000 LVCM food webs, each of which had a different initial cascade topology with different associated rates. Each LVCM was evolved dynamically in time until all possible extinctions occurred so that the reduced food web was in a stable equilibrium state.
Since it is clear from the results of Section \ref{sec:efficiency} that the maximum biological efficiency value of one does not ensure the persistence of food webs, we excluded the efficiency parameter and allowed the $a_{ij}$ and $a_{ji}$ rates to be randomly selected independent of each other. Furthermore, the birth rates are uniformly drawn from the interval $(r_b, r_b+1)$,  the death rates are uniformly distributed on the interval $(-r_d-1, -r_d)$,  the self-regulation rates are drawn uniformly from $(-r_r-1,-r_r)$, the $a_{ij}$ rates are chosen uniformly from $(-r_u-1,-r_u)$, and the $a_{ji}$ rates  are uniformly distributed on $(r_l, r_l+1)$. {Unit intervals are chosen to be consistent with previous work simulating LVCM dynamics\citep{chen2001}.} Here, for each of the 400,000 realizations, $r_b$, $r_d$, $r_r$, $r_u$, and $r_l$ are independently  drawn uniformly from $(0,30)$ {to allow for a greater variety of possible dynamics}.  For example, one could have the distributions given by $b_i\in (2.58,3.58)$, $d_i\in (-27.87,-26.87)$, $a_{ii}\in (-12.15,-11.15)$, $a_{ij}\in (-5.91,-4.91)$, and $a_{ji}\in (7.56,8.56)$. Each of these unit length distributions are chosen randomly for each of the 400,000 realizations.

To make the analysis more tractable, we reduced the dimension of the rate information by considering the absolute sums of the five types of rates. The absolute sum of birth rates, death rates, self-regulation rates, upper triangular rates, and lower triangular rates are given respectively as
\begin{equation}
        B = \sum_{1\le i\le S}{|b_i|},\label{e:B} 
\end{equation}
\begin{equation}
     D = \sum_{1\le i\le S}{|d_i|}, 
\end{equation}
\begin{equation}
      R = \sum_{1\le i\le S}{|a_{ii}|}, 
\end{equation}
\begin{equation}
       U = \sum_{1\le i<j\le S}{|a_{ij}|}, 
\end{equation}
\begin{equation}
        L = \sum_{1\le j<i\le S}{|a_{ij}|}.\label{e:L}
\end{equation}

Table \ref{tab:table1} shows the absolute sums of the five types of rates (Equations (\ref{e:B})-(\ref{e:L})) for five of the 400,000 realizations. The last column of the table shows the number of extinctions, $E$, which occurred when these five LVCM food webs were dynamically evolved in time. Note that these five example realizations show a wide range of extinctions, including one instance when the original 50 species cascade food web was able to persist with no extinctions. 

Given the values of $B$, $D$, $R$, $L$, and $U$ for a particular realization, it is natural to consider how their relative sizes affect the number of extinctions which occur when the LVCM food web is evolved dynamically. One can arrange the five absolute sums
 in ascending order based on their magnitudes for each realization. In doing so, one arrives at an inequality where $\alpha < \beta < \gamma < \delta < \epsilon$. Here, $\alpha, \beta, \gamma, \delta, \epsilon \in \{B, D, R, L, U\}$, and there are 120 possible orderings of the inequality. The last column of Table \ref{tab:table1} shows the corresponding inequalities for the five displayed realizations. 
\begin{table}[t!]
  \centering
\begin{tabular}{|c|c|c|c|c|c|c|}
\hline
$B$&$D$ & $R$ &$L$  &  $U$ & $E$ &$\alpha < \beta < \gamma < \delta < \epsilon$\\
\hline
327.48&87.41 & 999.89 & 1705.64  &  55.28  & 0 & $U<D<B<R<L$\\
\hline
189.39& 313.68& 352.99& 1408.26& 321.70& 15 & $B<D<U<R<L$\\
\hline
371.04& 913.40& 152.59& 1613.21& 306.21& 22& $R<U<B<D<L$\\
\hline
162.37& 27.80& 369.56& 2205.23& 2387.59& 30&$D<B<R<L<U$\\
\hline
200.59& 1144.64& 911.97& 472.31& 1881.74& 40& $B<L<R<D<U$\\
\hline
\end{tabular}
  \caption{Absolute sums of the rates (Equations (\ref{e:B})-(\ref{e:L})) for five example realizations of the LVCM. The number of extinctions for the realization is shown in the $E$ column. The corresponding inequality is shown in the last column.}
  \label{tab:table1}
\end{table}
For each inequality $\alpha < \beta < \gamma < \delta < \epsilon$ (of which there are 120), we form a cluster containing all realizations which satisfy the inequality.
For each cluster $C$, we can form the extinction frequency vector $E_C = (e_0, \ldots, e_{50})$ where $e_i$ represents the number of realizations in cluster $C$ with $i$ extinctions. Given such a vector $E_C$, the sum of entries, $\sum e_i$ tells us the size of cluster $C$. Similarly, $(\sum ie_i)/(\sum e_i)$ tells us the expected number of extinctions in the cluster. 
Notably, only seven clusters contained realizations which persisted with no extinctions, i.e., clusters with $e_0 > 0$. The vast bulk of these are found in clusters $(D< U< B< R< L)$ and $(U<D<B<R<L)$, each of which have 63 realizations which persist with zero extinctions, i.e., $e_0 = 63$.
It is worth noting 
that in both of the inequalities associated with these two clusters, the sum of the lower triangular interaction rates exhibits the highest absolute magnitude of all the rate sums. {Appendix A contains a table (Table \ref{tab:Cineqs}) denoting for each cluster number, $C$, the inequality, the number of extinctions, $e_i$, for the minimum value of $i$, and the number of realizations associated with each cluster}.

To better understand how the rates affect the extinction number, we consider the 120 clusters according to $E_C$ in reverse lexicographic order. More formally, let $E_C = (e_0,\ldots, e_{50})$ and let $E_{C'} = (e_0',\ldots, e_{50}')$ where $C$ and $C'$ represent two clusters.  We say $E_C<E_{C'}$ if $e_i > e_i'$ in the first position $i$ where $E_C$ and $E_{C'}$ differ. We assign cluster numbers $1,2,\ldots, 120$ based on this ordering. So for example cluster numbers 1, 2 and 3 correspond to the inequalities $(U<D<B<R<L)$, $(D< U< B< R< L)$ and $(D<U<B<L<R)$ since their corresponding extinction frequency vectors are $(63, 82, 90,\ldots)$, $(63, 78, 105, \ldots)$ and $(10, 33, 46, \ldots)$. 

 \begin{figure}[t!]
\centering\includegraphics[scale=0.45]{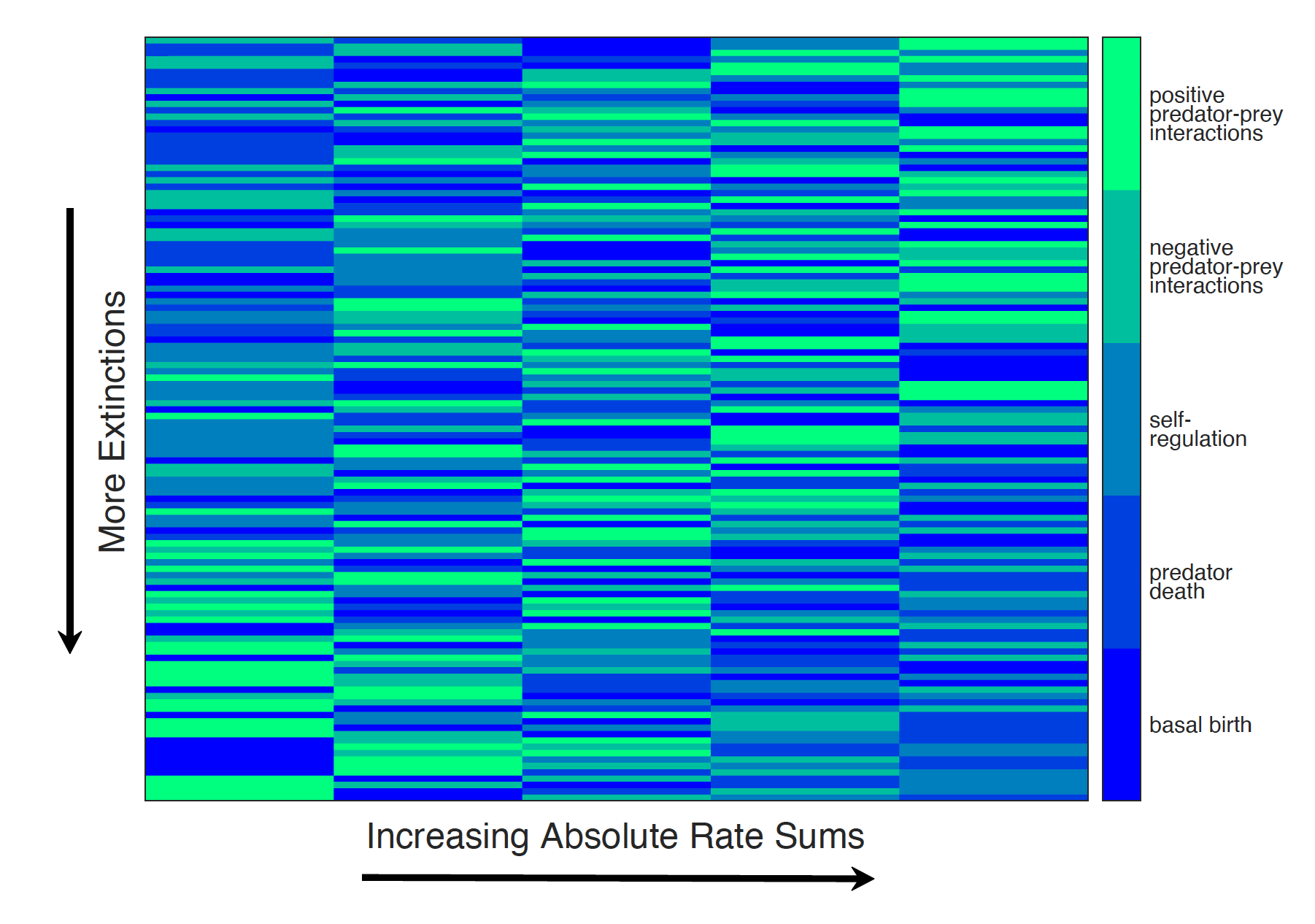}
\caption{Ordering of the 120 clusters according to the number of extinctions occurring in each cluster. Each row represents an inequality by displaying the order of the absolute rate sums, each of which is associated with a specific color.  Clusters with fewer extinctions are located at the top of the figure, and as one descends, the clusters have an increasing number of extinctions. 
}
\label{fig:order}
\end{figure} 
Figure \ref{fig:order} is a visualization of the extinction ordering of the 120 clusters. The first row is associated with cluster 
$C=1$, the second row with cluster $C=2$, and so on until one arrives at the last row which is associated with cluster $C=120$. 
The columns of the figure
denote the inequality associated with each cluster (row) by associating a specific color with each of the five absolute rate sums. Figure \ref{fig:order} shows that, in general, food webs with fewer extinctions (top rows) are characterized by an inequality which transitions from blue to green colors. On the other hand, food webs with more extinctions (bottom rows) are characterized by an inequality which transitions from green to blue colors. 

It is noteworthy that clusters in the top part of the figure are often associated with inequalities with positive interaction rate sums having the largest magnitude of all the rate sums. And when this is not the case, it is still often true that the magnitude of the positive rate sum is greater than the negative interaction rate sum.  Although we eliminated the efficiency parameter, $e$, relating the positive and negative interactions, one can use the efficiency relation $a_{ji}=-ea_{ij}$ to compute an efficiency proxy, $e_p$, given by 
\begin{equation}
    e_p=\frac{\sum |a_{ji}|}{\sum |a_{ij}|} =\frac{L}{U}.\label{e:ep}
\end{equation}
Equation (\ref{e:ep}) enables one to clearly see that inequalities/clusters which give rise to zero or only a few extinctions have a high efficiency proxy greater than one. Although mathematically one can consider any value of efficiency, values greater than one make no biological sense.

\begin{figure}[t!]
\centering\includegraphics[scale=0.65]{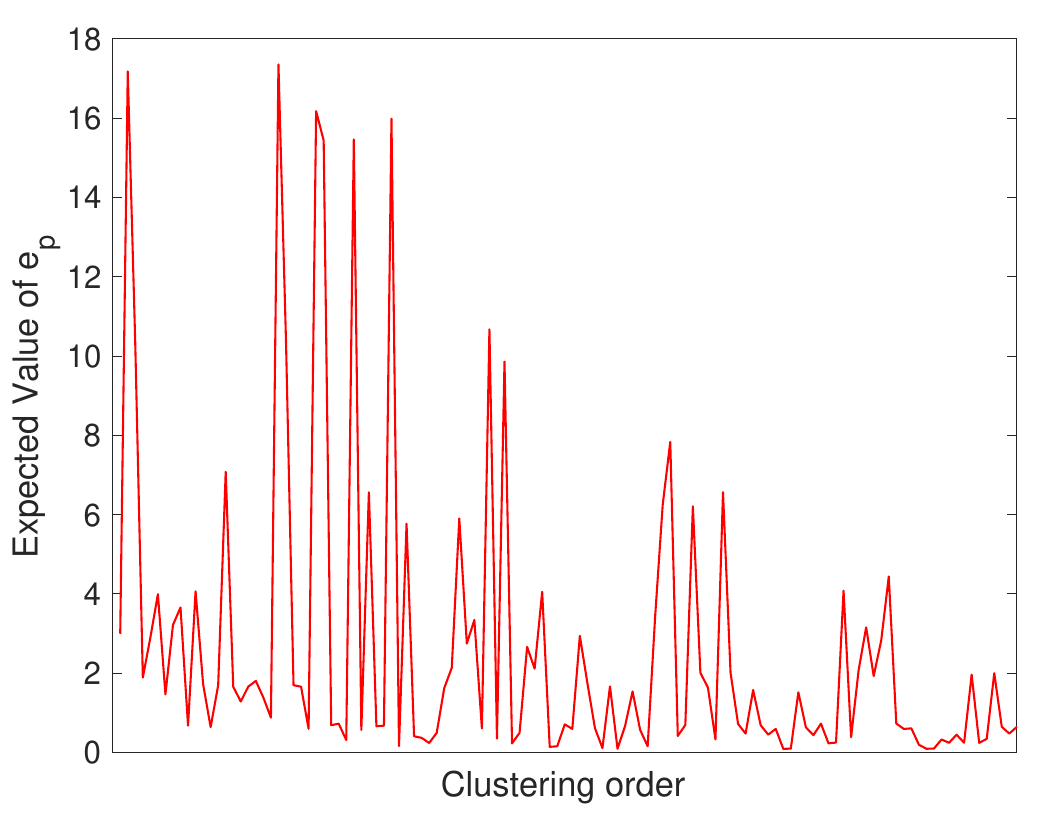}
\caption{Expected value of the efficiency proxy, $e_p$, for each of the 120 clusters. The cluster ordering is the same as was used in Figure \ref{fig:order}.
}
\label{fig:order_eff}
\end{figure}

Additionally, for each cluster (ordered as shown in Figure \ref{fig:order}), we calculated $e_p$  for every realization in the cluster and computed the expected value. The results are shown in Figure \ref{fig:order_eff}. Consistent with the previous discussion, one sees high $e_p$ values much larger than one for clusters residing in the the top rows of Figure \ref{fig:order}. One also sees a descending trend in the efficiency proxy so that clusters with large numbers of extinctions have a much lower value of $e_p$. There is variance in Figure \ref{fig:order_eff} because the birth and death rates also play a significant role in determining the number of extinctions, but they are not included in the computation of $e_p$. The importance of the birth and death rates will be expanded upon in the following section.
 
As we have seen, LVCM food webs with an associated inequality/cluster in the top rows of Figure \ref{fig:order} will typically have fewer extinctions than LVCM food webs with an associated inequality/cluster in the middle and bottom rows of Figure \ref{fig:order}. Therefore, if we generate a LVCM food web associated with a middle or bottom row cluster, we will typically observe many extinctions as the system is evolved dynamically in time. The resulting reduced food web is a new stable food web at equilibrium (e.g. Figures \ref{fig:cascade} and \ref{fig:cascade.L-v}).  Importantly, the new stable food web may no longer possess the required structure to make it a cascade food web.

Since the new food web is at a stable, coexistence equilibrium, if one were to start with this food web and evolve it according to the same rates, no further extinctions would occur. Therefore, one might hypothesize that the inequalities/clusters associated with the reduced, stable food webs would reside in the upper rows of Figure \ref{fig:order}. As an example, the LVCM shown in Figure \ref{fig:cascade} is associated with cluster $C=28$ ($B<D<R<U<L$), while the reduced, stable food web shown in Figure \ref{fig:cascade.L-v} is associated with cluster $C=8$ ($D<U<L<B<R$). Cluster $C=8$ is not one of the seven clusters which exhibit zero extinctions for the cascade food webs. This is due to the fact that the reduced food webs are not necessarily cascade food webs. Nevertheless, cluster $C=8$ resides much higher in Figure \ref{fig:order} than cluster $C=28$, and as such satisfies the hypothesis.

To get a more quantitative sense, we generated 1000 LVCM food webs which were evolved to the reduced, stable food web at equilibrium. Dividing the clusters into groups of 30 (1-30, 31-60, 61-90, 91-120), we found that the reduced food webs had associated clusters which more likely fell in the first 2 groups as opposed to the latter 2 groups. Specifically, 27 of the first set of 30 clusters and 24 of the second set of 30 clusters were achieved, often with high frequency. In contrast, only 15 of the third set of clusters and 2 of the fourth set of clusters were achieved, and many of these were achieved only a single time. There is clearly a bias toward the reduced stable food webs being associated with a cluster in the upper part of Figure \ref{fig:order}.

\subsection{\label{sec:prediction-c}Prediction of the Number of Extinctions}
It is reasonable to think that one can use the clustering results of Section \ref{sec:persistence} as a simple predictive model. Specifically, given a new LVCM food web, one can determine the absolute rate sums and the inequality associated with them. Then, rather than evolving the system dynamically to see how many extinctions occur, one can predict the number of extinctions using the expected value of extinction number for the cluster associated with the specific inequality.

\begin{figure*}
\begin{minipage}{.5\textwidth}
\includegraphics[scale=0.41]{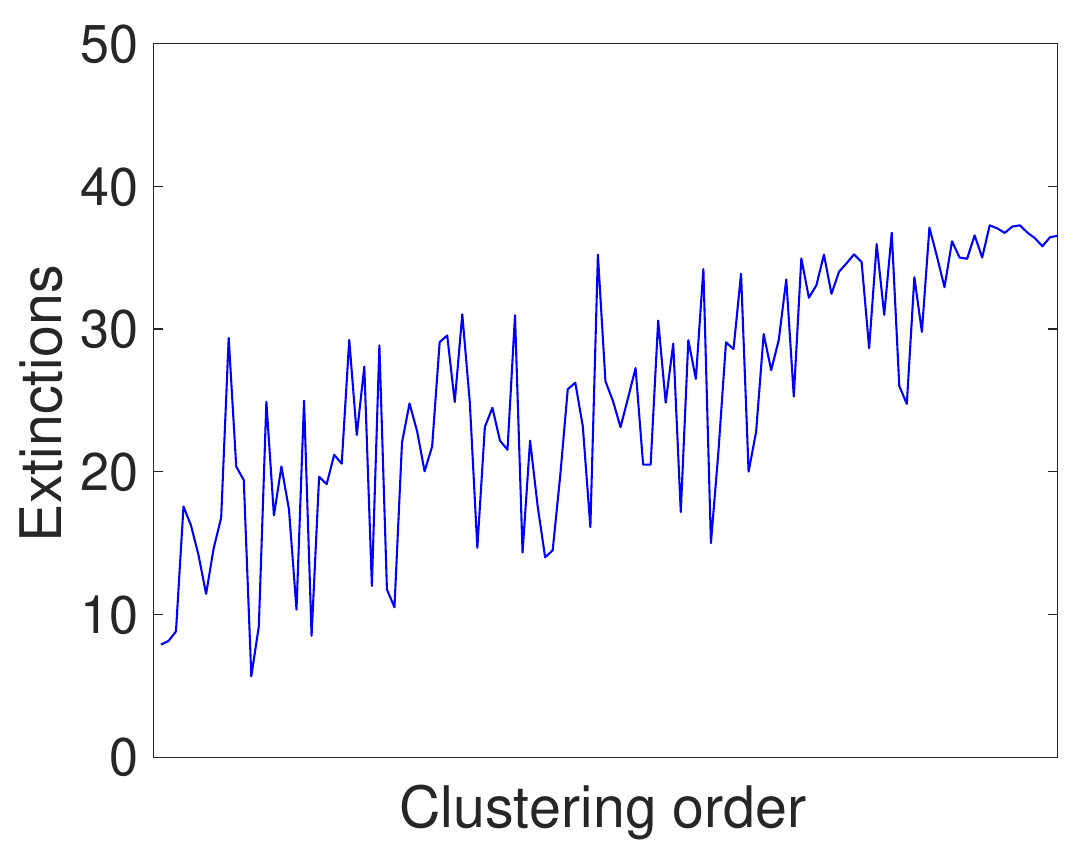}
\end{minipage}%
\begin{minipage}{.5\textwidth}
\includegraphics[scale=0.41]{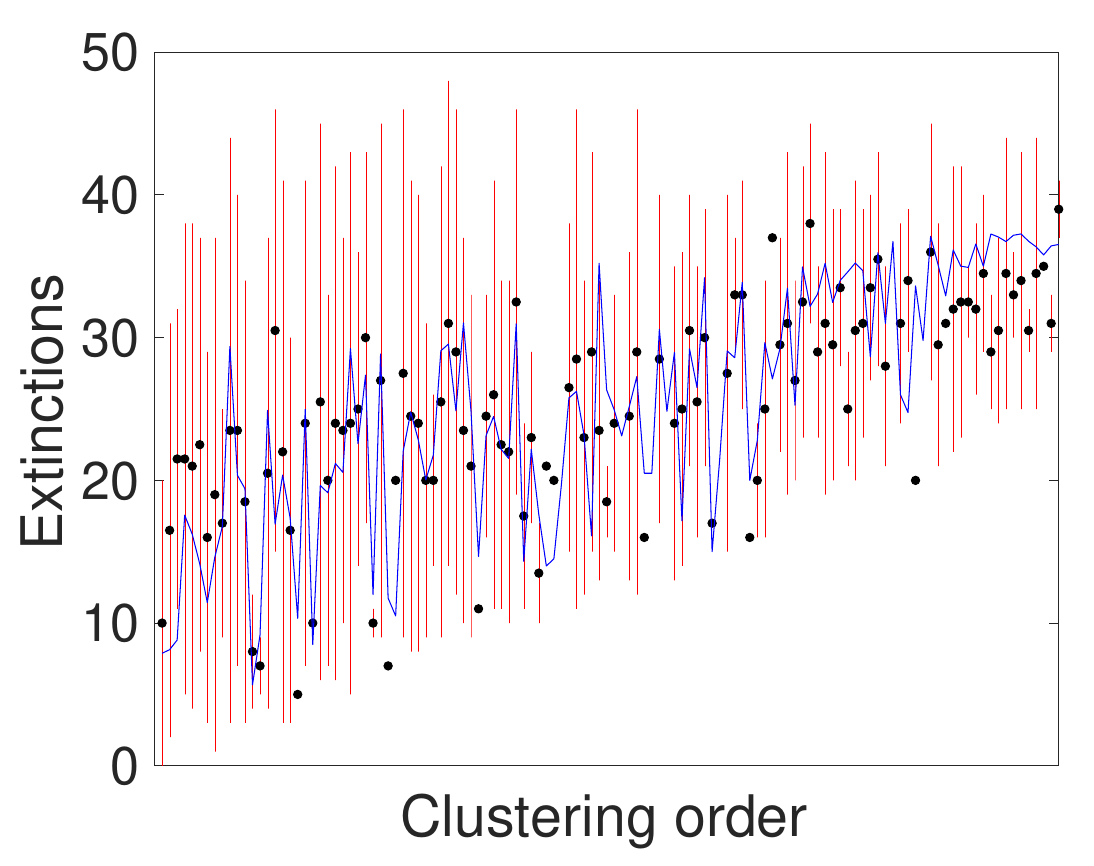}
\end{minipage} 
\caption{(a) Expected value of extinctions for each cluster in the training set. (b) Expected value of extinctions for each cluster in the training set (blue curve) and test set (black dots). The red bars denote the minimum and maximum number of extinctions in each cluster of the test set. The cluster ordering is the same as was used in Figure \ref{fig:order}. }
\label{fig:cluster prediction EXT.no}
\end{figure*}

To explore how well this simple predictive model works, we divided the 400,000 realizations into training and test sets, allocating 80\% of the realizations for training and 20\% of the realizations for validation. Using the cluster ordering of Figure \ref{fig:order}, Figure \ref{fig:cluster prediction EXT.no}(a) shows the expected number of extinctions in each cluster. Consistent with the results of Section \ref{sec:persistence}, one sees an increase in the expected number of extinctions as one moves to the right. 
These expected values are used to predict the number of extinctions for a new LVCM food web, thus reducing significantly the computational time to find the number of extinctions by actually evolving the system dynamically in time.

To assess the accuracy of the model's predictive performance, we computed the expected number of extinctions for each cluster in the test set. The values are shown as black dots in Figure \ref{fig:cluster prediction EXT.no}(b), and one sees they compare favorably with the predicted expected values shown in blue (the blue curves in Figure \ref{fig:cluster prediction EXT.no}(a) and (b) are identical). In fact, by the Law of Large Numbers, as the number of realizations in the test set increases, the expected values of the test set clusters will approach the predicted expected values. However, there is also a large variance in the possible number of extinctions occurring amongst all the realizations in each cluster. The red bars in Figure \ref{fig:cluster prediction EXT.no}(b) denote the minimum and maximum number of extinctions observed in each cluster of the test set. Due to the large variance, this simple predictive model may not perform as well for individual realizations.

To check this, we calculated the coefficient of determination, $R^2$, using
\begin{equation}
R^2=1-\frac{\sum (y_{i}-\hat{y}_{i})^2}{\sum (y_{i}-\bar{y})^2}\label{equ:R^2},
\end{equation}
where $y_i$ represents an individual realization in some cluster in the test set, $\bar{y}$ is the expected value of the test set cluster within which the realization resides, and $\hat{y}$ indicates the predicted expected value found using the appropriate cluster in the training set data. The result, $R^2=0.48$, is, because of the variance, quite poor. Therefore, to improve our predictive capabilities, we employed the random forest algorithm \citep{breiman2001}. 

 We constructed the random forest with the open-source data analytics application Radiant \citep{Radiant}, which uses R statistical software \citep{R}. Specifically, our random forest consisted of 20 decision trees, which were created through the training process. The Radiant package creates each tree based on a  a random piece of the training data, with a different random piece of the data used for each tree. Due to the random selection of training data, each tree will have some data points that were not used for the training. These unused data points are referred to as out-of-bag samples. Once the training is complete, these out-of-bag samples which have not been seen by the decision trees can be used to check the predictive performance of each decision tree.

The random forest produced an $R^2=0.87$, which is much better at predicting the number of extinctions compared with the simple cluster-based approach described above. Although the random forest acts as a ``black-box", one can still leverage the model to assess the relative importance of the five types of rates based on their contribution to predicting the number of extinctions. 

We used a permutation feature importance approach in which the model's performance was measured when the values of the features (absolute rate sums) are randomly shuffled. Specifically, after training, the values of each of the five features are permuted in the out-of-bag samples, and the error is computed for the perturbed set of data. Then, the importance score of each feature is found by averaging over all the decision trees the difference in the out-of-bag error before permutation and after permutation. Lastly, the score is normalized by the standard deviation of the differences. Features with larger values for the importance score are considered to be more important than features with smaller values \citep{zhu2015}.
Table \ref{tab:importance} shows the order of importance and accompanying importance scores. 

\begin{table}[h!]
  \centering
\begin{tabular}{|c|c|c|c|c|}
\hline
$B$& $D$& $L$& $R$& $U$ \\
\hline
0.66&0.62 & 0.45 & 0.26  & 0.22 \\
\hline
\end{tabular}
\caption{Importance order and scores of the five absolute rate sums. }
  \label{tab:importance}
\end{table}

It is clear that the absolute sum of the birth and death rates, $B$ and $D$, have the most substantial impact on the prediction of species extinction. The absolute sum of the rates below the diagonal, $L$, are also important, consistent with the role of efficiency proxy given by Equation (\ref{e:ep}) and discussed in Section \ref{sec:persistence}. The absolute sum of the diagonal rates and rates above the diagonal, $R$ and $U$, are less important.
\subsection{\label{sec:prediction}Prediction of the Number of Extinctions: A Neural Network Approach}
\begin{figure*}
\centering\includegraphics[scale=0.9]{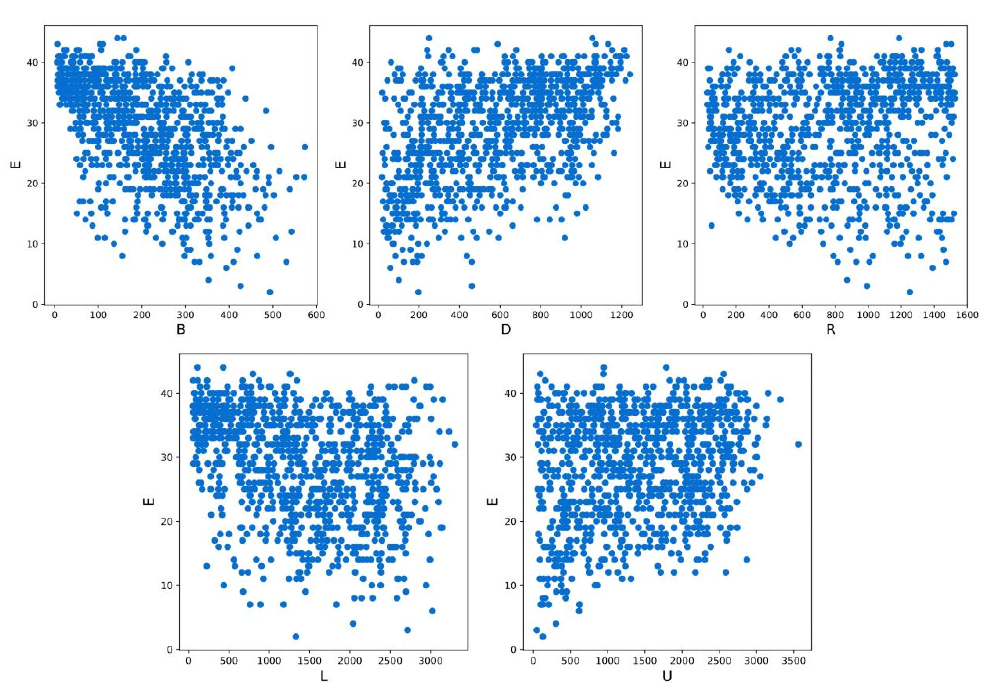}
\caption{Correlation between the explanatory variables, $B$, $D$, $R$, $L$, and $U$, and the response variable, $E$.}
\label{fig:corr-rates}
\end{figure*}

\begin{figure*}[t!]
\centering
\includegraphics[scale=0.3]{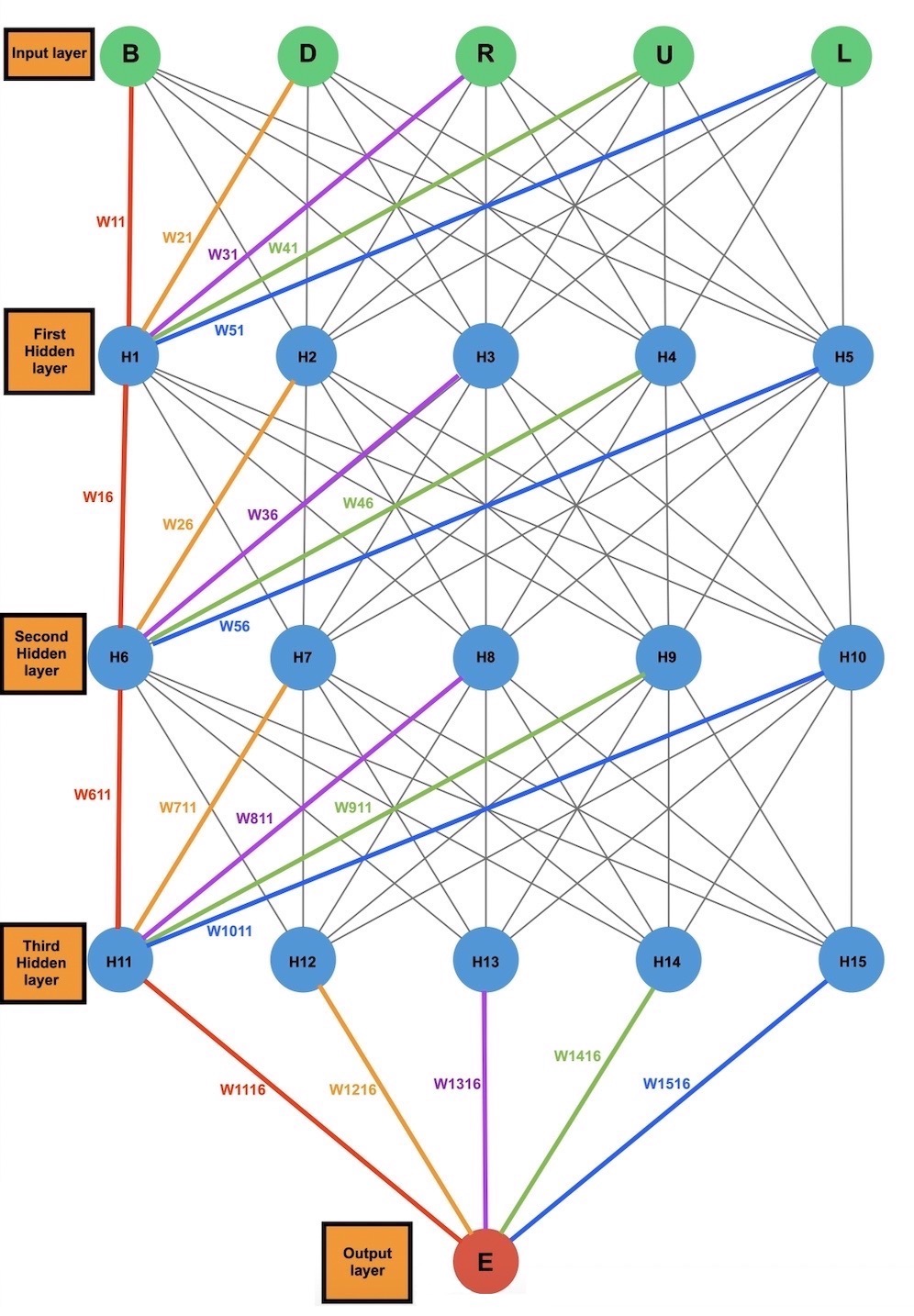}
\caption{An artificial neural network (ANN) with three hidden layers, each of which contains five nodes. The input layer incorporates the five absolute rate sums, $B$, $D$, $R$, $L$, and $U$. The $H_i$ denote the nodes in the hidden layers, and the $W_{ij}$ denotes the weight values in each layer, only some of which are labeled. The output layer predicts the number of extinct species, $E$.}
\label{fig:NN}
\end{figure*}

\begin{figure}[t!]
\centering\includegraphics[scale=0.4]{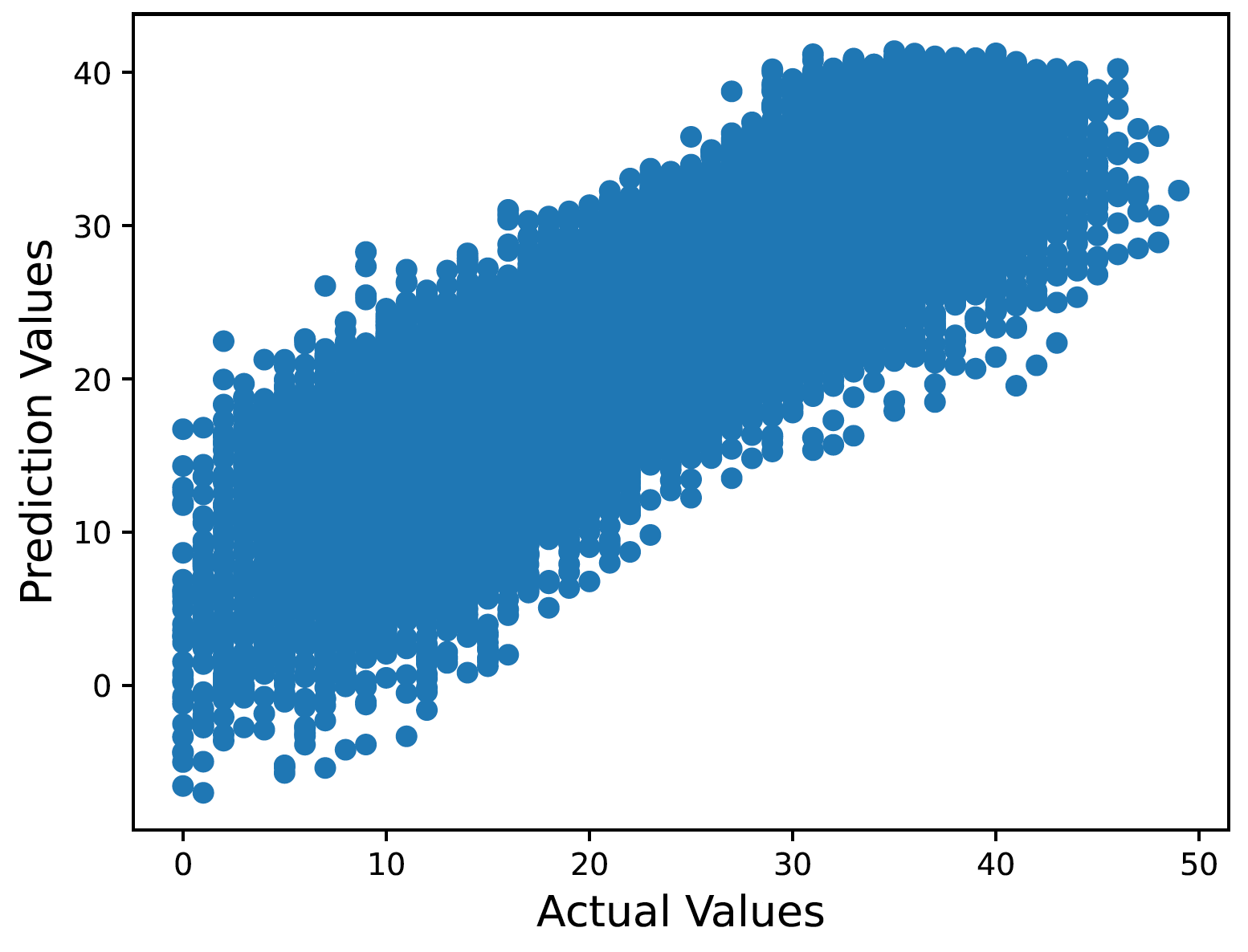}
\caption{Predicted number versus the actual number of extinctions}
\label{fig:pred-actual}
\end{figure}

In Section \ref{sec:prediction-c} we saw that the cluster-based approach was inadequate for predicting the number of extinct species. In contrast, the random forest approach provided much more accurate predictions along with a global estimation of feature importance found by averaging the effect of permutations over many realizations. We now endeavor to employ an artificial neural network model to provide accurate predictions while simultaneously enabling insight into the predictive method for individual realizations. In this neural network approach, the absolute rate sums, $B$, $D$, $R$, $L$, and $U$ serve as input or explanatory variables, while the output or response variable is the number of extinctions for a particular realization, $E$. As with the clustering and random forest approaches, the objective is to predict the number of species in an LVCM food web that will go extinct without having to explicitly simulate the complex food web dynamics in time.\\ 
Figure \ref{fig:corr-rates} shows the relationship between each of the explanatory variables and the response variable. Each filled circle represents a data point associated with a random simulation of the LVCM. While all the plots show good correlation, the absolute sums of the birth and death rates exhibit the strongest correlations. This is consistent with the importance order and scores of the five absolute rate sums presented in Section \ref{sec:prediction-c}. Taken together, the correlation plots of Figure \ref{fig:corr-rates} provide strong justification that a neural network approach incorporating all five explanatory variables can be used to accurately predict the number of extinct species.
We used an artificial neural network (ANN) with three hidden layers, each of which contains five nodes (see Figure \ref{fig:NN}). We employed a linear, or identity, activation function for the first and second hidden layers, and a $\tanh{x}= \frac{e^x-e^{-x}}{e^x+e^{-x}}$ activation function for the third hidden layer. The dataset of 400,000 LVCM realizations was partitioned into training (80\%) and testing (20\%) sets. During the training, we used 50 epochs and the adam optimizer, with loss computed according to the mean squared error. We computed the coefficient of determination, $R^2=0.81$, by averaging the $R^2$ values for five independent training runs using the neural network architecture shown in Figure \ref{fig:NN} and described above. Figure \ref{fig:pred-actual} shows the high accuracy of the predicted number of extinctions compared with the actual number of extinctions.

Typically in machine learning, one uses nonlinear activation functions for each of the hidden layers. However, doing so prevents one from obtaining an associated equation which can be used to provide insight into the predictive method. Therefore, we used linear activation functions for the first two layers. Moreover, we did explore the use of nonlinear activation functions for all the hidden layers, and the performance was not quite as good as what we achieved with the set-up described above. Separately, because the linear activation function provides a linear combination of the input variables, theoretically one should be able to collapse the linear hidden layers into a single hidden layer. In practice, however, we found that using two hidden layers with the linear activation function provides better predictive performance (average $R^2=0.815$) than using a single hidden layer with the linear activation function (average $R^2=0.72$) or three (average $R^2=0.811$) or four (average $R^2=0.80$) hidden layers with the linear activation function (all scenarios had a final hidden layer using the $\tanh{x}$ activation function).

To derive the predictive equation, one can compute the value of each node $H_i$ shown in Figure \ref{fig:NN} by finding the linear combinations associated with the weights $W_{ij}$ and the preceding nodes serving as inputs. To find the values of $H_i$ ($i=1\ldots 5$) in the first hidden layer, form the linear combinations involving the input variables $B$, $D$, $R$, $L$, and $U$, and the weights $W_{ij}$, $i,j=1\ldots 5$ so that 
\begin{equation}
H_i = B\cdot W_{1i} + D\cdot W_{2i} + R\cdot W_{3i} + U\cdot W_{4i} + L\cdot W_{5i} +c_i \label{e:H1}
\end{equation}
where $c_i$ is a constant bias term.
Similarly, the values of $H_i$ ($i=6,\ldots,10$) in the second hidden layer are found as 
\begin{equation}
    H_i = H_{1}\cdot W_{1i} + H_{2}\cdot W_{2i} + H_{3}\cdot W_{3i} + H_{4}\cdot W_{4i} + H_{5}\cdot W_{5i} +c_i
\end{equation}
and the values of the nodes in the third hidden layer before the $\tanh{x}$ activation function is applied, $X_i$ ($i=11,\ldots,15$), are found as 
\begin{equation}
X_i = H_{6}\cdot W_{6i} + H_{7}\cdot W_{7i} + H_{8}\cdot W_{8i} + H_{9}\cdot W_{9i} + H_{10}\cdot W_{10i} +c_i
 \end{equation}
After applying the activation function, one has
 \begin{equation}
H_i=\frac{e^{X_i}-e^{-X_i}}{e^{X_i}+e^{-X_i}},\quad i=11,\ldots,15.
 \end{equation}
 Lastly, the equation to predict the number of extinctions is formed as
 \begin{equation}
E=\sum\limits_{i=11}^{15} H_i\cdot W_{i16} +c_{16}.\label{e:Hl}
\end{equation}
Using our neural network architecture and Equations (\ref{e:H1})-(\ref{e:Hl}), we can find the specific predictive equation for one training run (associated with an $R^2=0.82$). The $X_i$, $i=11,\ldots,15$, and $E$ equations are given as
\begin{equation}
\begin{split}
    X_{11} &= 1.26095 + 0.0117789 B - 0.00164158 D \\
    &\qquad + 0.000457352 L - 0.000837343 R - 0.00030985 U,
\end{split}
\end{equation}
\begin{equation}
\begin{split}
    X_{12} &= 0.184674 + 0.00257395 B - 0.00116786 D\\
    &\qquad + 0.000767944 L- 0.00103138 R - 0.00013743 U,
\end{split}
\end{equation}
\begin{equation}
\begin{split}
    X_{13} &= 1.0082 - 0.000260263 B + 0.00220154 D \\
    &\qquad - 0.0000475026 L- 0.000752652 R + 0.000240999 U,
\end{split}
\end{equation}
\begin{equation}
\begin{split}
    X_{14} &= -0.640453 - 0.0004638 B - 0.000523884 D \\
    &\qquad + 0.000552778 L- 0.00111785 R + 0.000170356 U,
\end{split}
\end{equation}
\begin{equation}
\begin{split}
    X_{15} &= -0.303342 - 0.01249 B + 0.000643048 D  \\
    &\qquad + 0.0000154759 L+ 0.000711079 R + 0.0000790523 U
\end{split}
\end{equation}
\begin{equation}
\begin{split}
    E &= 17.431087 + 10.999312 \tanh{(X_{11})} - 14.078264 \tanh{(X_{12})} \\
    &\quad + 16.728575 \tanh{(X_{13})} + 8.828843 \tanh{(X_{14})}\\ &\quad+ 9.922335 \tanh{(X_{15})}.\label{equ:prediction}
\end{split}
\end{equation}
Figure \ref{fig:pred-actual} shows very good agreement between the predicted number of extinctions found using Equation (\ref{equ:prediction}) versus the actual number of extinctions for the 80,000 realizations contained within the test set.

In order to gain an indication of the importance of the respective inputs, as we did with the random forest, we drew on a gradiant approach, a method associated with the growing repertoire of interpretability of neural networks \citep{samek2021}. Because Equation (\ref{equ:prediction}) is relatively simple with limited nonlinearity, the function can be approximated locally by a linear function given as
\begin{equation}
    E({\bf x})\approx \sum\limits_{i=1}^5 \frac{\partial E}{\partial {\bf x}_{(i)}}({\bf x}_0)({\bf x}_{(i)}-{\bf x}_{0_{(i)}})=\sum\limits_{i=1}^5 R_i, \label{e:approx}
\end{equation}
where ${\bf x}=(B,D,L,R,U)$ and ${\bf x}_0$ is a nearby root. Each of the $R_i$ terms can be interpreted as the contribution of feature $i$ (i.e., $B$, $D$, $L$, $R$, $U$) to the prediction of extinction number $E$. 

Each of the partial derivatives in Equation (\ref{e:approx}) can be easily computed. Given a specific realization of $B$, $D$, $L$, $R$, $U$ values, one can use standard numerical root finder methods to find a nearby root. For example, consider a realization from a cluster which resides near the top of Figure \ref{fig:order}. The absolute rate sums associated with this realization are $(B, D, L, R, U) = (359.69, 60.42, 1595.20, 1309.81, 97.51)$. The actual number of extinctions associated with this realization is zero, while the predicted number using Equation (\ref{equ:prediction}) is 0.76. Using these values as an initial guess, a nearby root is given by $(B, D, L, R, U) = (366.54, 47.30, 1596.68, 1312.13, 95.93)$. Evaluating the partial derivatives at this root point, Equation (\ref{e:approx}) becomes
\begin{equation}
\begin{split}
E\approx&-0.0223*(359.69-366.54)\\
    &+0.0435*(60.42-47.30)\\
    &-0.0048*(1595.20-1596.68)\\
    &-0.0080*(1309.81-1312.13)\\
    &+0.0053*(97.51-95.93).
\end{split}
\end{equation}
Therefore, the contribution of each feature, $R_i$ is given as  
$$R_D = 0.57 > R_B= 0.15 > R_R= 0.02 > R_U= 0.008 > R_L= 0.007.$$ By comparison with Table \ref{tab:importance} one can see that the feature contribution here is different from the importance order found using the random forest. It is important to remember that the random forest result is an average over many thousands of realizations while this local linearization result is specific to one particular realization. If one were to consider a second realization with absolute rate sums near the root point given as $(B, D, L, R, U) = (226.61, 117.49, 2201.20, 1497.13, 108.40)$, and where the predicted number of extinctions is 2.40 versus the actual number of zero, then one finds that the feature contributions, $R_B=3.11$, $R_D = 3.05$, $R_L=-2.88$, $R_R=-1.47$, and $R_U= 0.07$, match the random forest importance ordering. While the random forest provides a global average of feature contributions, the local linearization approach provides information to help explain the predictions of specific realizations.

The predictive equation for number of extinctions given by Equation (\ref{equ:prediction}) was determined using LVCM food webs containing 50 species. The result can also be used for LVCM food webs containing different numbers of species due to linear scaling.  We now demonstrate that cascade food webs have an expected number of basal and non-basal species which scale linearly with the number of species in the food web. Similarly, we will show that the interaction strengths also scale linearly with the number of species. Therefore the magnitude of the absolute rate sums, $B$, $D$, $R$, $L$, and $U$, all scale linearly according to the number of the species. Because of this, one can scale the absolute rate sums for $S$ species by $50/S$ so that the rate sums are on an equivalent 50 species scale. Then one can use Equation (\ref{equ:prediction}) to predict the number of extinctions on a 50 species scale. Finally, by scaling this number of extinctions by $S/50$, one has a prediction of number of extinctions for the $S$ species food web.   

To find the expected number of basal species,
consider a cascade model food web $G$ with $S$ species labeled from $[S]:=\{ 1,2,\ldots, S\}$, with connectance probability $p$. By construction, all edges are directed from higher index species to lower indexed species. Every connection of the form $ij$, where $i>j$, appears independently with probability $p$. Basal species are those with an out-degree of zero.  We first compute the expectation of $n_b$, the number of basal species. For a given species $i\in\{1,\ldots, S\}$, the probability that $i$ has out-degree zero is given by $(1-p)^{i-1}$ since all edges from $i$ to the set $[i-1]$ must not be present. Thus, by linearity of expectation, the expected number of basal species is
\begin{equation}
\E{n_b} = \sum_{i=1}^S (1-p)^{i-1} \frac{1-(1-p)^{S+1}}{1-(1-p)} = \frac{1-(1-p)^{S+1}}{p}.
\end{equation}
For our simulations, we use $p=c/S$ with $c=3.72$ \citep{cohen1985stochastic}. In this case, we have that the expected number of basal species is asymptotically $\frac{1-e^{-c}}{c}S \approx .2623 S$. Since the expected number of basal species grows linearly in $S$, the expected number of non-basal species will also grow linearly in $S$. Therefore the absolute sum of birth and death rates, $B$ and $D$, will grow linearly in $S$.

Now let $I$ represent the number of interactions (edges) in $G$. For our value of $p = c/S$ with $c=3.72$, we have
\begin{equation}
\mathbb{E}[I] = \binom{S}{2}\cdot p  \sim \frac{c}{2}S  = 1.86S.
\end{equation}
Since the number of interactions scales linearly in $S$, the absolute sum of the interaction rates, $L$ and $U$, will scale linearly in $S$. Finally, since all species have a self-regulation rate, clearly, the absolute sum of self-regulation, $R$, scales linearly in $S$. 

\begin{center}  
\begin{figure}[h]
\centering\includegraphics[scale=0.65]{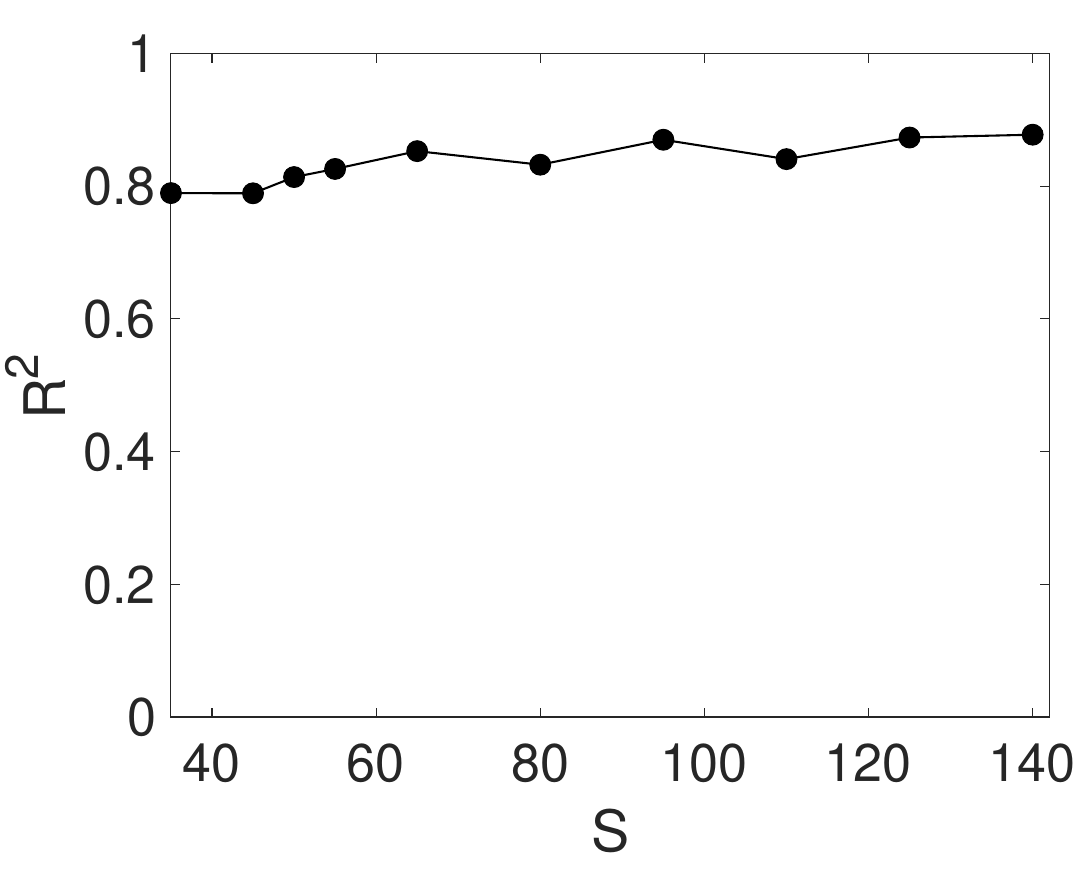}
\caption{Coefficient of determination, $R^2$, comparing the predicted versus actual number of extinctions for a selection of LVCM food webs containing different numbers of species, $S$. 
}
\label{fig:r^2-N}
\end{figure} 
\end{center}

We considered LVCM food webs of size $S \in \{35, 45, 55, 65, 80, 95, 110, 125, 140\}$. For each value of $S$, we generated 500 LVCM realizations using the rate distributions as described in Section \ref{sec:persistence} ($r=30$). As described previously, the absolute rate sums for each realization were scaled by $50/S$, and  Equation (\ref{equ:prediction}) was used to predict the number of extinctions for an equivalent 50 species food web. These values were then scaled by $S/50$ to obtain a prediction for the appropriate $S$-sized food web. Figure \ref{fig:r^2-N} shows the $R^2$ values for 10 different values of $S$. One can see excellent predictive capability for a wide range of $S$ values. 

\subsection{Effect of Death Rates on Extinction Order}\label{sec:order}
We have shown that LVCM food webs exhibit a substantial number of species extinctions and that the birth and death rates, $B$ and $D$ have the most substantial impact. To further explore the importance of these rates, we measured the sum of the birth/death rates of surviving species as a function of the number of extinctions. For the simulations performed in this section, we return to coupling the interaction rates via efficiency, and use $e=0.1$ 
and $r=1$ for the rate distributions. 

\begin{figure}[h!]
\includegraphics[scale=0.8]{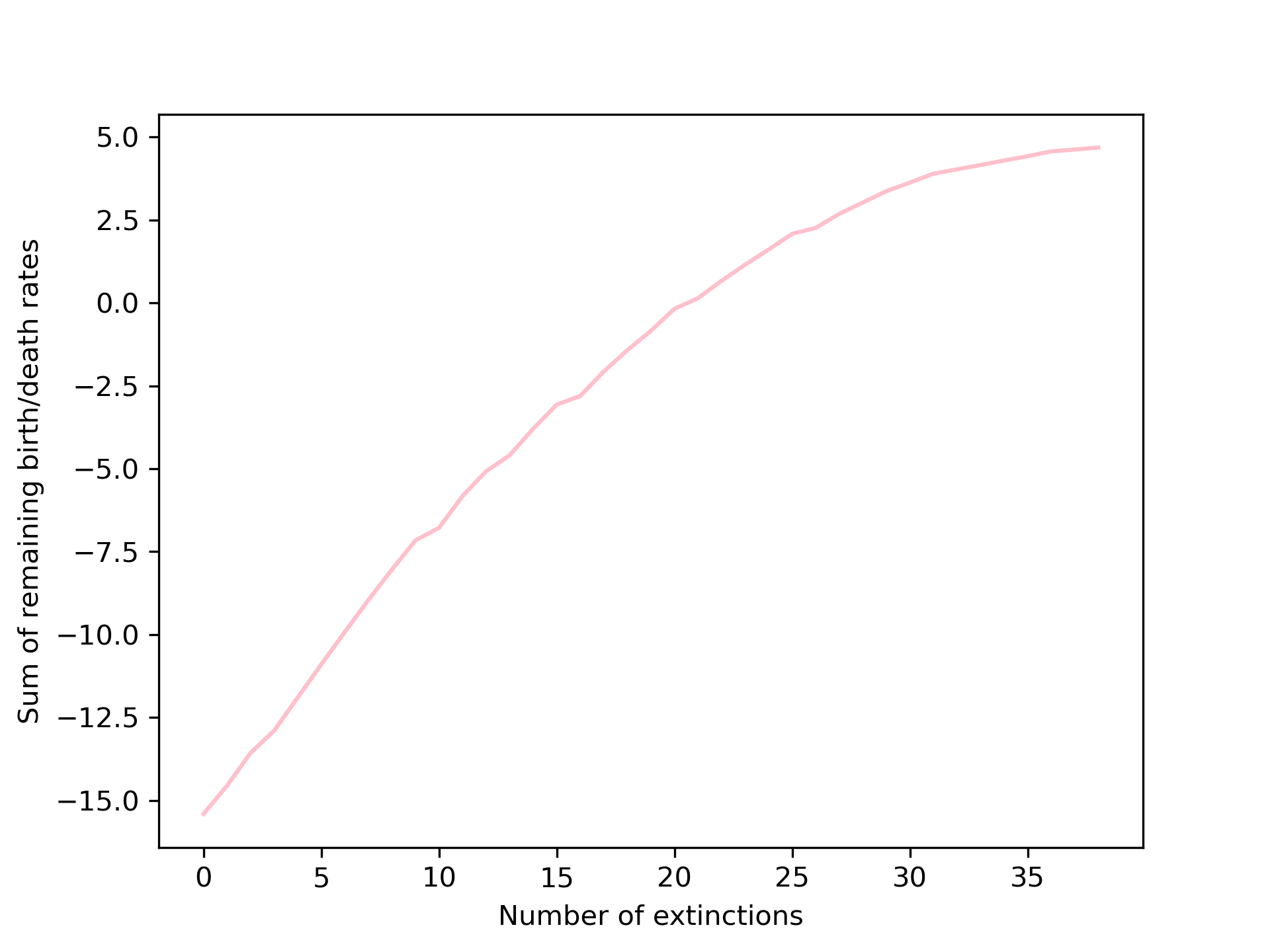}
\caption{Sum of the birth/death rates of the surviving species as a
function of the number of extinctions for a 50 species LVCM food web.}
\label{fig:death-ext-single}
\end{figure}

Figure \ref{fig:death-ext-single} shows this relationship for one realisation of a 50 species food web and a choice of rates. At the start, for zero extinct species, the sum of the birth/death rates is negative since there are far more non-basal species with negative growth (death) rates than basal species with positive growth (birth) rates. As the number of extinctions increases, the sum of the rates becomes less negative, and then becomes positive. Eventually, as the food web evolves to a stable food web at equilibrium, the sum of the rates reaches a plateau.

The result was compared with the corresponding sum for a simulation where extinctions occurred at random rather than according to the Lotka-Volterra dynamics. By comparing the two outcomes for different rate sums, we determined that the death rates of the species have the most influence in determining which species will go extinct as well as the order of extinction. Although the result may seem trivial, due to the complicated interplay of dynamics it is not at all clear a priori that this would be the case. 

\begin{figure}[h!]
\centering
\includegraphics[scale=0.8]{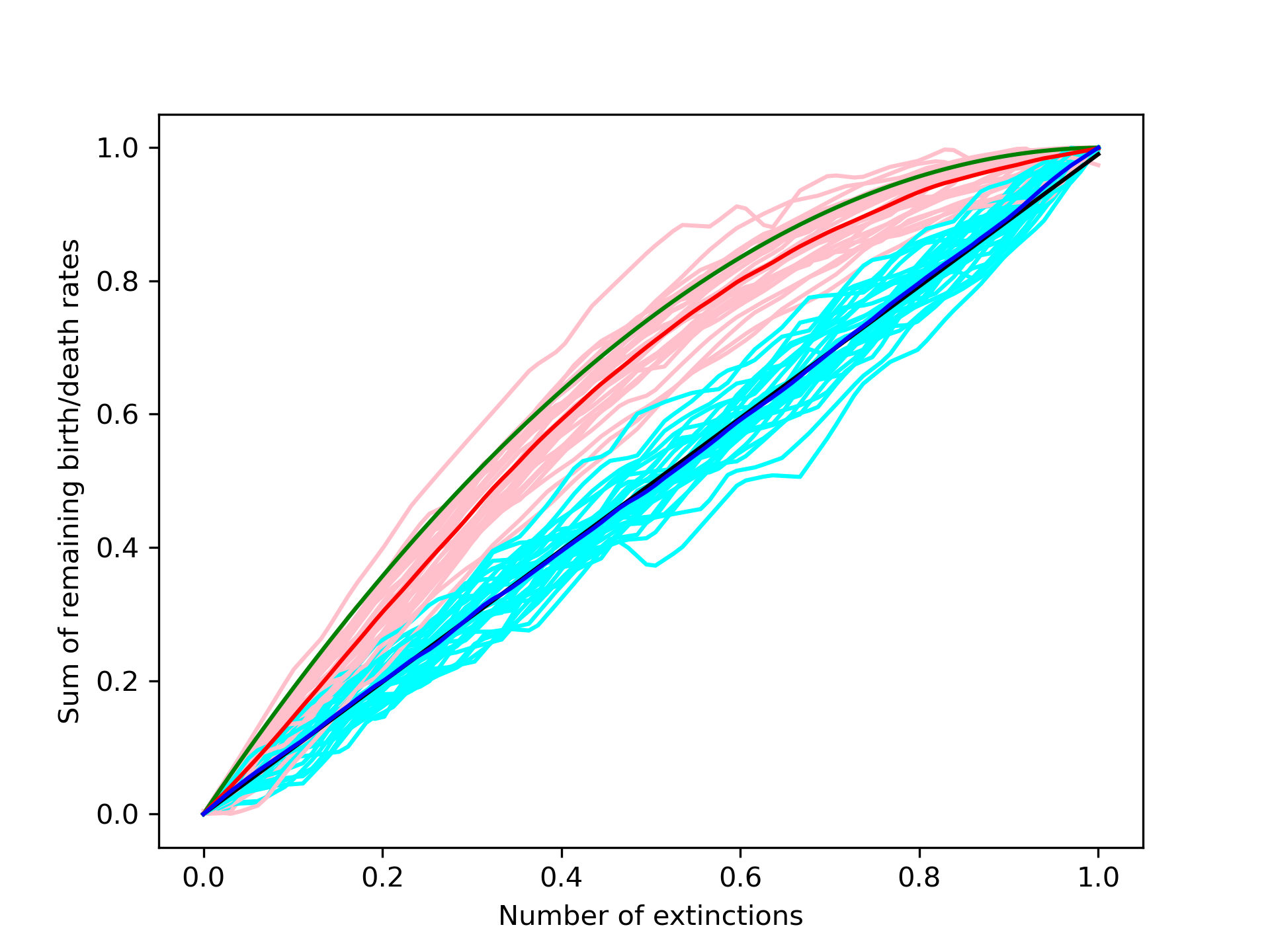}
\caption{Normalized sum of the birth/death rates of the surviving species as a function of the normalized number of extinctions for 50 species cascade food webs. The pink curves show the normalized sum of the combined birth and death rates for the species that remain in the food web for 30 realizations of the LVCM. The red curve depicts the average of the 30 pink curves. The cyan curves show the normalized sum of the birth/death rates for the same cascade food webs, but with extinctions occurring randomly. The single dark blue curve depicts the average of the 30 cyan curves. The green and black curves are the theoretical predictions for the Lotka-Volterra and random extinction cases, respectively.}
\label{fig:death-ext}
\end{figure}

To perform this comparison, we would like to consider many realizations of food web and choice of rates. Because each realization can have differing numbers of extinct species as well as a different range for the sum of the birth/death rates, we must normalize both the abscissa and ordinate values to lie between zero and one.
Figure \ref{fig:death-ext} shows the combined normalized sum of the birth rates (for basal species) and death rates (for non-basal species) as a function of the normalized number of extinctions for cascade food webs of 50 species with two types of dynamical interactions.

The first type of interactions are due to Lotka-Volterra dynamics per the LVCM described above. As before, the cascade food webs are evolved in time according to the Lotka-Volterra dynamics, and a number of species extinctions are observed until eventually a new stable food web is achieved. With each species extinction, we calculated the combined sum of the birth and death rates of the species that remain in the food web.  This process of summing the birth and death rates was repeated until the food web reached its stable, equilibrium configurations. In Figure \ref{fig:death-ext}, the pink curves show the sum of the combined birth and death rates of all remaining species in the food web as a function of extinction number for 30 realizations of the LVCM (different food web topologies and different rates in the Lotka-Volterra dynamics). As noted above, because each realization may have different sum values and different numbers of extinction, we normalized both to the unit interval. The single red curve represents the average of these 30 different realizations.

We compare these results by considering the exact same cascade food webs, but instead of extinctions occurring due to Lotka-Volterra dynamics, now species are randomly chosen for extinction.  As the species go extinct, we perform the same procedure as for the Lotka-Volterra dynamics, and sum the combined birth and death rates for the surviving species. The cyan curves in Figure \ref{fig:death-ext} show the normalized results for  thirty realizations. The single dark blue curve represents the average of these thirty scenarios. 

There is a notable difference between the cyan/dark blue curves and the pink/red curves. The pink curves are larger than the cyan curves, and  the slope of the pink curves is larger than the slope of the cyan curves at the beginning of the process. It is therefore evident that under Lotka-Volterra dynamics, there is a biased behavior in the extinction process, whereby species with higher death rates go extinct first on average.

The computational results can be confirmed theoretically. First we derive the theoretical curve associated with random extinction events. Consider the growth rates (birth and death) associated with the species which go extinct (in their extinction order), and which we denote as $b_1,\ldots, b_n$. Let  $B = b_1 + \ldots+ b_n$, 
and let $S_i = b_{i+1} + \ldots + b_n + A$, where $A$ is the sum of the birth/death rates for the surviving species. Therefore,
$S_i$ is the partial sum for all but the first i terms, and is what is shown by the pink curve in the pre-normalized sum shown in Figure \ref{fig:death-ext-single}.  The curve values range from $B+A$ to $A$ as $i$ ranges from $0$ to $n$.

Now let $\sigma$ be a random permutation of $n$. Then $b_\sigma(1), b_\sigma(2),\ldots , b_\sigma(n)$ are the birth/death rates in a random order.
The expectation of every $b_\sigma(i)$ is the average of the $b_i$ terms, and is given by $B/n$. We let $b= B/n$.
Therefore, the expectation of the randomized partial sum $b_\sigma(i+1),\ldots, b_\sigma(n) + A$, is $(n-i)b + A = B+A - ib$.

The graph of this expectation of the randomized partial sums from $0$ to $n$ is a line with slope given by $-b$. In Figure \ref{fig:death-ext}, the graph is given by the black line, which agrees very well with the dark blue curve found computationally by averaging 30 realizations. Note that the slope $-b$ appears positive in the graph because the $b_i$ growth rates are mostly negative death rates associated with non-basal species. 
Importantly, this argument has nothing to do with the distribution of the birth rates. No matter what values are ascribed to the rates $b_1,\dots ,b_n$, if one randomizes their order, the partial sums will be linear in expectation.

Now we derive the theoretical curve associated with the Lotka-Volterra dynamics. The pink curves in Figure \ref{fig:death-ext} suggest that species with higher death rates go extinct first. To derive the theoretical results, we will assume  
 that non-basal species will in fact go extinct in the order of their death rates. Note that if all initial conditions were the same, and there were no interactions between species, then this would be the case.
 
Suppose that there are $N$ non-basal species and that their death rates are uniformly distributed on the interval $[-a,-b]$. Then the $k$th lowest (most negative) death rate has expected value $-a + \frac{k}{N+1}(a-b)$ (see for example \cite{order_stat}).
Let $D_j$ represent the expected sum of the remaining death rates after $j$ extinctions.
Then by linearity of expectation, $D_j$ is the sum of all the expected death rates except for the $j$ most negative ones. In other words, 
\begin{equation}
\begin{split}
D_j &= \sum_{i=j+1}^N  -a + \frac{i}{N+1}(a-b) \\
&= -a(N - j) + \frac{a-b}{N+1}\of{\frac{N(N+1)}{2} - \frac{j(j+1)}{2}}.
\end{split}
\end{equation}
If one lets $x=j/N$, then 
\begin{equation}
D_j = N\cdot\of{-a(1-x) + \frac{a-b}{2}(1-x^2)} + \mathcal{O}(1).
\end{equation}
Note that as the width of the interval increases, the quadratic part of the term is ``scaled up''. 
After normalization to the unit square, we get \[d(x) = 1- (1-x)^2,\] which appears as the green curve in Figure \ref{fig:death-ext}.

The green theoretical curve in Figure \ref{fig:death-ext} has the same shape as the red curve, which was found by averaging 30 realizations. However, although the two curves are close in value, they are not perfectly overlaid with each other. This is due to the fact that the theoretical curve was derived based on an assumption that all species went extinct according to their death rate. While the very close agreement between curves suggests that the species under Lotka-Volterra dynamics generally go extinct according to their death rates, there will be some exceptions. Nevertheless, it is clear that the death rate is a significant driver of extinction as well as the order of extinction.
\section{\label{sec:conc}Conclusion}
This work considers the synthesis of cascade food webs with nonlinear Lotka-Volterra equations to better understand the role of population dynamics and trophic structure in ecological communities. Importantly, the Lotka-Volterra equations incorporate a biological efficiency. This efficiency, which is widespread in mathematical ecology, links the predator-prey interactions. Through simulation, we have shown how the cascade topology coupled with a biological efficiency leads to numerous species extinctions. This suffocating quality of efficiency was also found in another common synthetic food web model, the niche model. This has potentially far-reaching consequences for the assumptions underlying not only synthetic models, but in the way community matrices might be built from empirical data.

Moreover, we showed via clustering analysis that persistence could be achieved only when the absolute sums of the birth, death, self-regulation, and interaction rates satisfied specific inequalities. Even then, only a very small proportion of LVCM food webs persisted intact when the dynamics were evolved in time. Importantly, we showed using an efficiency proxy that these persistent LVCM food webs are not biologically realistic. 

With only the simplified rate sums of the variables, we explored the use of
machine learning to predict the number of extinctions that would occur in a given LVCM food web without actually evolving the dynamics. Both a random forest approach, and an artificial neural network, were highly effective in predicting extinctions, which also provides insight into how the structure of the values comprising the system plays the most crucial role in the viability of the underlying dynamics.
Lastly, we explored the internal processes involved in the extinction of species during the unfolding dynamics of a system, and have been able to highlight how the death rates play the dominant role in determining the species' extinction order.

Returning to the point made above, the work presented in this article leads to an important cautionary message. When field ecologists or mathematical ecologists use linearized community matrices as their starting point, without considering the dynamics that would have led to such a linearized system, then certain assumptions may have been made which would make such a Jacobian unlikely to have been generated. While the process of reverse engineering the underlying dynamics from a community matrix is not generally feasible due to loss of information and non-uniqueness, it is perhaps a reasonable suggestion that any assumptions hard-wired into a linearized system be applied at the dynamical systems level to investigate how such assumptions have an effect on systems as they unfold in time.

\clearpage

\section*{Declarations}

\subsection*{Funding}
This work was funded by the National Science Foundation (Award Nos. DMS-1853610 and CNS-1625636).

\subsection*{Conflicts of interest/Competing Interests}

The authors have no relevant financial or non-financial interests to disclose.

\subsection*{Authors' Contributions}
Deepak Bal, Michael A.S. Thorne, and Eric Forgoston conceived and developed the study. Sepideh Vafaie developed the machine learning algorithms. All authors were involved in the theoretical and computational analysis. Eric Forgoston secured funding and managed the project. All authors wrote, commented on, and edited the manuscript. All authors read and approved the final draft of the manuscript.

\clearpage

\clearpage
\appendix*
\section{Comprehensive Cluster Overview}
\begin{table*}[!h]
    \centering
    \begin{minipage}[b]{0.32\linewidth}
        \centering
        \tiny
        \begin{tabular}{|c|c|c|c|}
            \toprule
            $C$ & $\alpha < \beta < \gamma < \delta < \epsilon$ & $e_{min(i)}$ & $\sum{e_i}$ \\
            \midrule
            1&$U<D<B<R<L$&$e_0=63$&1,514\\
            \hline
            2&	$D<U<B<R<L$	&$e_0=63$	&1,766\\
            \hline
            3&$D<U<B<L<R$&$e_0=10$&	657\\
            \hline
            4	&$D<B<U<R<L$&	$e_0=8$&	8,004\\
            \hline
            5	&$U<B<D<R<L$&	$e_0=5$&	6,010\\
            \hline
            6&	$U<D<B<L<R$&	$e_0=4$&	534\\
            \hline
            7&$D<B<U<L<R$	&$e_0=1$&	2,108\\
            \hline
            8&	$D<U<L<B<R$&	$e_1=2$&	110\\
            \hline
            9&	$U<D<R<B<L$&	$e_1=1$&	154\\
            \hline
            10&	$B<U<D<R<L$&	$e_2=2$&	15,924\\
            \hline
            11&	$U<B<R<D<L$&	$e_2=1$&	3,431\\
            \hline
            12&	$D<L<U<B<R$&	$e_2=1$&	116\\
            \hline
            13&	$U<D<L<R<B$&	$e_2=1$&	6\\
             \hline
             14	&$D<U<R<L<B$&	$e_2=1$&	8\\
             \hline
             15	&$B<D<U<R<L$&	$e_3=4$	&19,624\\
             \hline
             16&	$D<B<R<U<L$	&$e_3=2$&	13,671\\
             \hline
             17&	$D<B<L<U<R$&	$e_3=1$&	2,092\\
             \hline
             18	&$D<U<R<B<L$&	$e_3=1$&	200\\
             \hline
             19&	$D<U<L<R<B$	&$e_3=1$&	10\\
             \hline
             20&	$D<L<B<U<R$	&$e_3=1$&	630\\
             \hline
             21&	$U<D<R<L<B$&	$e_3=1$	&5\\
             \hline
             22&	$D<B<R<L<U$	&$e_4=1$	&13,627\\
             \hline
             23&	$U<R<D<B<L$&	$e_4=1$	&156\\
             \hline
             24	&$D<B<L<R<U$&	$e_5=1$	&7,931\\
             \hline
             25	&$U<R<B<D<L$	&$e_5=1$&	1,067\\
             \hline
             26&	$U<B<D<L<R$&	$e_5=1$&	1,699\\
             \hline
             27&	$U<D<L<B<R$&	$e_6=2$&	88\\
             \hline
             28&	$B<D<R<U<L$&	$e_6=1$	&39,925\\
             \hline
             29&	$D<L<U<R<B$&	$e_7=3$&	9\\
             \hline
           30	&$B<U<R<D<L$&	$e_7=2$	&8,400\\
           \hline
           31	&$U<R<D<L<B$&	$e_7=2$	&6\\
           \hline
           32	&$U<R<L<D<B$&	$e_7=1$&	3\\
           \hline
           33	&$D<R<B<U<L$&	$e_8=4$&	2,532\\
           \hline
           34	&$D<L<B<R<U$&	$e_8=2$&	1,780\\
           \hline
           35	&$D<R<B<L<U$&	$e_8=2$&	2,437\\
           \hline
           36	&$D<R<U<B<L$	&$e_8=1$	&232\\
           \hline
           37	&$U<R<B<L<D$	&$e_8=1$&	196\\
           \hline
           38	&$B<R<U<D<L$&	$e_9=2$	&8,344\\
           \hline
           39	&$B<R<D<U<L$&	$e_9=2$	&26,931\\
           \hline
           40	&$R<D<B<U<L$	&$e_9=1$	&2,247\\
            \bottomrule
        \end{tabular}
    \end{minipage}%
    \hfill
    \begin{minipage}[b]{0.32\linewidth}
        \centering
        \tiny
        \begin{tabular}{|c|c|c|c|}
            \toprule
            $C$ & $\alpha < \beta < \gamma < \delta < \epsilon$ & $e_{min(i)}$ & $\sum{e_i}$ \\
            \midrule
           41	&$B<D<U<L<R$&	$e_9=1$	&4,382\\
           \hline
           42	&$R<L<D<B<U$	&$e_9=1$	&189\\
           \hline
           43	&$D<L<R<U<B$	&$e_9=1$&	11\\
           \hline
           44	&$R<U<D<B<L$&	$e_9=1$	&196\\
           \hline
           45	&$R<U<B<D<L$&	$e_{10}=4$	&1,265\\
           \hline
            46 &$D<R<L<B<U$& $e_{10}=2$ & 219\\
           \hline
           47	&$D<L<R<B<U$&	$e_{10}=1$	&212\\
           \hline
           48	&$B<D<R<L<U$	&$e_{10}=1$	&39,756\\
           \hline
           49	&$R<U<D<L<B$&	$e_{10}=1$&	7\\
           \hline
           50	&$R<U<L<B<D$&	$e_{10}=1$&	61\\
           \hline
           51	&$R<D<U<L<B$&	$e_{10}=1$	&7	\\
           \hline
           52	&$U<L<R<D<B$	&$e_{10}=1$	&3\\
           \hline
           53	&$R<D<L<U<B$	&$e_{10}=1$&	6\\
           \hline
           54	&$L<D<R<U<B$&	$e_{10}=1$	&10\\
           \hline
           55	&$R<B<U<D<L	$&$e_{11}=4$	&3,945\\
           \hline
           56	&$R<B<D<U<L$	&$e_{11}=3$	&10,504\\
           \hline
           57	&$R<D<U<B<L$	&$e_{11}=2$	&199\\
           \hline
           58	&$U<L<D<R<B$&	$e_{11}=1$&	9\\
           \hline
           59	&$B<U<D<L<R	$&$e_{11}=1$	&3,912\\
           \hline
           60	&$L<D<R<B<U$&	$e_{11}=1$	&183\\
           61	&$R<D<L<B<U$	&$e_{11}=1$	&196\\
           \hline
           62	&$R<U<B<L<D$&	$e_{11}=1$&	246\\
           \hline
           63	&$R<D<B<L<U$&	$e_{12}=1$&	2,253\\
           \hline
           64	&$R<B<D<L<U$&$e_{12}=1$	&10,445\\
           \hline
           65	&$R<L<D<U<B$&$e_{12}=1$&	11\\
           \hline
           66	&$R<L<U<D<B$&	$e_{12}=1$&	6\\
           \hline
           67	&$B<R<D<L<U$	&$e_{13}=2$&	27,427\\
           \hline
           68	&$U<R<L<B<D$&	$e_{13}=2$&	45\\
           \hline
           69	&$U<B<R<L<D$&	$e_{13}=1$&	528\\
           \hline
           70&	$R<U<L<D<B$&	$e_{13}=1$&	6\\
           \hline
           71&	$R<L<B<D<U$&	$e_{14}=2$&	1,134\\
           \hline
           72&	$R<B<U<L<D$&	$e_{14}=1$&	598\\
           \hline
           73&	$B<D<L<U<R$&	$e_{14}=1$&	4,492\\
           \hline
           74&	$D<R<U<L<B$&	$e_{14}=1$&	3\\
           \hline
           75&	$L<R<D<U<B$&	$e_{14}=1$&	7\\
           \hline
           76&	$R<B<L<D<U$&	$e_{15}=4$&	3,947\\
           \hline
           77&	$R<L<B<U<D$&	$e_{15}=1$&	226\\
           \hline
           78&	$B<D<L<R<U$&	$e_{16}=3$&	19,314\\
           \hline
           79&	$D<R<L<U<B$&	$e_{16}=2$&	3\\
           \hline
           80&	$L<R<U<D<B$&	$e_{16}=1$&	7\\
            \bottomrule
        \end{tabular}
    \end{minipage}%
    \hfill
    \begin{minipage}[b]{0.32\linewidth}
        \centering
        \tiny
        \begin{tabular}{|c|c|c|c|}
            \toprule
            $C$ & $\alpha < \beta < \gamma < \delta < \epsilon$ & $e_{min(i)}$ & $\sum{e_i}$ \\
            \midrule
           81&	$U<L<D<B<R$&	$e_{16}=1$&	84\\
           \hline
           82&	$L<R<D<B<U$&	$e_{17}=1$&	193\\
           \hline
           83&	$R<B<L<U<D$&	$e_{17}=1$&	618\\
           \hline
           84&	$L<D<B<R<U$&	$e_{17}=1$&	1,467\\
           \hline
           85&	$R<L<U<B<D$&	$e_{17}=1$&	57\\
           \hline
           86&	$U<L<B<R<D$&	$e_{17}=1$&	219\\
           \hline
           87&	$B<R<U<L<D$	&$e_{18}=2$	&1,081\\
           \hline
           88&	$L<R<B<D<U$&	$e_{18}=1$&	1,067\\
           \hline
           89&	$U<B<L<D<R$&	$e_{18}=1$&	1,253\\
           \hline
           90&	$L<D<U<B<R$	&$e_{18}=1$&	101\\
           \hline
           91&	$U<B<L<R<D$&	$e_{19}=2$&	537\\
           \hline
           92	&$L<D<B<U<R$	&$e_{18}=1$	&516\\
           \hline
           93	&$B<R<L<D<U$	&$e_{19}=3$	&8,261\\
           \hline
           94	&$B<U<R<L<D$	&$e_{19}=2$	&1,057\\
           \hline
           95	&$U<L<R<B<D$	&$e_{19}=1$	&57\\
           \hline
           96	&$L<B<R<D<U$&	$e_{20}=1$&	3,435\\
           \hline
           97	&$L<R<U<B<D$&	$e_{20}=1$&	56\\
           \hline
           98	&$B<L<R<D<U$&	$e_{20}=1$&	8,445\\
           \hline
           99	&$L<U<R<D<B$&	$e_{20}=1$&	5\\
            \hline
           100	&$L<D<U<R<B$&	$e_{20}=1$&	4\\
            \hline
            101	&$L<U<R<B<D$	&$e_{20}=1$&	88\\
            \hline
            102	&$L<U<D<R<B$	&$e_{20}=1$	&6\\
            \hline
            103&$B<L<D<R<U$&	$e_{21}=1$&	16,085\\
            \hline
            104&	$U<L<B<D<R$	&$e_{21}=1$	&426\\
            \hline
            105&	$L<U<R<B<D$&	$e_{22}=2$&	44\\
            \hline
            106	&$L<B<D<R<U$&	$e_{22}=1$	&6,005\\
            \hline
            107&	$B<R<L<U<D$&	$e_{23}=3$	&1,104\\
            \hline
            108&	$L<R<B<U<D$&	$e_{23}=1$&	218\\
            \hline
            109&	$L<B<R<U<D$&	$e_{23}=1$&	502\\
            \hline
            110&	$L<U<B<R<D$&	$e_{24}=2$&	211\\
            \hline
            111	&$B<U<L<R<D$&	$e_{24}=1$&	1,114\\
            \hline
            112&	$B<L<U<D<R$&	$e_{24}=1$&	2,870\\
            \hline
            113	&$B<U<L<D<R$&	$e_{25}=5$&	2,769\\
            \hline
            114	&$B<L<R<U<D$&	$e_{25}=2$&	1,100\\
            \hline
            115	&$B<L<U<R<D$&	$e_{25}=2$&	1,077\\
            \hline
            116&	$B<L<D<U<R$&	$e_{25}=1$&	3,945\\
            \hline
            117	&$L<B<U<D<R$	&$e_{25}=1$&	1,233\\
            \hline
            118&	$L<U<B<D<R$&	$e_{25}=1$&	439\\
            \hline
            119&	$L<B<D<U<R$&	$e_{25}=1$&	1,590\\
            \hline
            120	&$L<B<U<R<D$&	$e_{28}=1$&	507\\
            \bottomrule
        \end{tabular}
    \end{minipage}
    \caption{Cluster number, $C$, and associated inequality, number of extinctions, $e_i$, for the minimum value of $i$, and number of realizations.}\label{tab:Cineqs}
\end{table*}

\clearpage

\bibliographystyle{chaos}
\bibliography{Refs}

\end{document}